\documentclass[12pt]{article}
\usepackage[margin=1in]{geometry}
\usepackage{amsmath,amssymb,amsfonts}
\usepackage{tikz}
\usetikzlibrary{shapes,arrows,positioning,fit,backgrounds,calc}
\usepackage[table]{xcolor}
\usepackage{tabularx}
\usepackage{booktabs}
\usepackage{array}
\usepackage{graphicx}
\usepackage{caption}
\usepackage{float}
\usepackage{stfloats}
\usepackage{authblk}
\usepackage{setspace}
\usepackage{natbib}
\usepackage{hyperref}

\usepackage[document]{ragged2e}
\usepackage{enumitem}
\usepackage{colortbl}
\usepackage{array}
\usepackage{multirow}
\usepackage{caption}
\usepackage{multicol}
\usepackage{centernot}
\usepackage{mdframed}
\usepackage{pgfplots}
\usepackage{textgreek}
\usepackage{siunitx}
\pgfplotsset{compat=1.17}

\usepackage{xr-hyper}
\usepackage{xr} 
\usepackage{hyperref}

\makeatletter
\newcommand*{\addFileDependency}[1]{
  \typeout{(#1)}
  \@addtofilelist{#1}
  \IfFileExists{#1}{}{\typeout{No file #1.}}
}
\makeatother

\newcommand*{\myexternaldocument}[1]{%
    \externaldocument{#1}%
    \addFileDependency{#1.tex}%
    \addFileDependency{#1.aux}%
}
\myexternaldocument{Supplement}

\DeclareMathOperator{\argmin}{argmin}

\title{Performance of weakly-supervised electronic health record-based phenotyping methods in rare-outcome settings}
\author[1,*]{Yunjing Hong, MS}
\author[2,1]{Jennifer C. Nelson, PhD}
\author[2,1,3]{Brian D. Williamson, PhD}
\affil[1]{University of Washington, United States of America}
\affil[2]{Kaiser Permanente Washington Health Research Institute, United States of America}
\affil[3]{Fred Hutchinson Cancer Center, United States of America}
\affil[*]{Corresponding author. Email: yunjing.hong@cuanschutz.edu.}
\date{}

\begin{document}

\maketitle
\begin{abstract}
Accurately identifying patients with specific medical conditions is a key challenge when using clinical data from electronic health records. Our objective was to comprehensively assess when weakly-supervised prediction methods, which use \textit{silver-standard labels} (proxy measures of the true outcome) rather than gold-standard true labels, perform well in rare-outcome settings like vaccine safety studies. We compared three methods (PheNorm, MAP, and sureLDA) that combine structured features and features derived from clinical text using natural language processing, through an extensive simulation study with data-generating mechanisms ranging from simple to complex, varying outcome rates, and varying degrees of informative  silver labels. We also considered using predicted probabilities to design a chart review validation study. No single method dominated the other across all prediction performance metrics. Probability-guided sampling selected a cohort enriched for patients with more mentions of important concepts in chart notes. SureLDA, the most complex of the three algorithms we considered, often performed well in simulations. Performance depended greatly on selected tuning parameters. Care should be taken when using weakly-supervised prediction methods in rare-outcome settings, particularly if the probabilities will be used in downstream analysis, but these methods can work well when silver labels are strong predictors of true outcomes. 

\textbf{Keywords:} machine learning; weak supervision; computable phenotyping; electronic health records; vaccine safety. 

\end{abstract}


\doublespacing
\section*{Introduction}

A persistent challenge when using clinical data from electronic health records (EHRs) for research purposes is accurately identifying patients with specific medical conditions, also known as phenotyping \citep{Pathak2013,Richesson2013a,Hripcsak2013,Tian2013,Denny2013, Miotto2016}. In EHR-based vaccine safety studies \citep{Davis2005,Lieu2007,McNeil2014} of serious but rare outcomes like anaphylaxis, traditional rule-based phenotypes that use International Classification of Disease (ICD) diagnosis codes are often inaccurate \citep{Walsh2013, Bhatt2021, Denny2010}, and resource-intensive gold standard outcomes obtained via manual chart review are infeasible to scale \citep{carrell2024}. Rule-based algorithms \citep{Hripcsak2013,upadhyaya2017,campbell2023} that rely only on structured EHR data may not generalize to new populations or complex conditions \citep{Liao2019, Ferte2021, Becker2023}, encouraging the incorporation of unstructured data from clinical notes using natural language processing (NLP) to improve performance \citep{Liao2019, Yan2024}. Supervised machine learning methods can also be used, but these require large amounts of gold-standard data for training and evaluation \citep{carrell2023,carrell2024, Zhang2019}. These limitations have motivated new approaches that incorporate NLP-derived features and use less (or no) gold-standard training data \citep{bach2017, yu2018, Zhang2019a, PivovarovRimma2015, DeFreitas2021, Miotto2016, Nogues2022}.

PheNorm \citep{yu2018}, MAP \citep{Liao2019} (Multimodal Automated Phenotyping), and sureLDA \citep{Ahuja2020} (Surrogate-guided Latent Dirichlet Allocation) are computable phenotyping approaches that can incorporate both structured and unstructured data and are weakly-supervised, using silver label proxies for the gold-standard outcome in model training \citep{bach2017}. Although designed for chronic conditions with richly documented visit histories in the EHR, PheNorm has also been shown to accurately predict acutely-occurring COVID-19 disease \citep{Smith2024}. Challenges remain when using these methods to identify acutely-occurring vaccine adverse events. In addition to being assessed within a narrow post-vaccination follow-up window, these events are also rare \citep{Babazadeh2019}, yielding data sparsity. Further, event documentation in the EHR is inconsistent, as serious events typically occur in emergency departments or hospitals rather than in integrated outpatient settings where documentation practices are more standardized \citep{lai2022, campbell2023}.  Existing simulation studies that evaluate the performance of sureLDA and PheNorm \citep{Ahuja2020} do not consider a very rare vaccine outcome setting and do not evaluate the estimated probabilities of outcome against the true probabilities, which is important if these will be used in downstream analyses \citep{Xu2021}. Critical gaps remain regarding the minimum silver label quality required for adequate prediction performance in rare disease setting.

We conducted a comprehensive simulation study to evaluate the performance of PheNorm, MAP, and sureLDA in rare event settings where the ground truth was known. We varied the complexity of the data-generating mechanism, the outcome rate, and the informativeness of silver labels. We assessed both the ability to distinguish cases from controls and the accuracy of predicted probabilities. We also used an example dataset to demonstrate how predicted probabilities can guide chart selection for algorithm validation, ensuring balanced representation of cases and controls. This probability-guided approach more efficiently uses limited chart review resources compared to simple random sampling, which may yields insufficient cases in rare-outcome settings \citep{carrell2024}.

\section*{Methods}
\subsection*{Silver-standard Labels}
A distinguishing feature of weakly-supervised methods \citep{bach2017}, including PheNorm, MAP, and sureLDA, is their reliance on \textit{silver-standard labels}---proxy measures of the outcome derived directly from EHR data without manual expert review (we will also refer to these as \textit{silver labels}). Examples include the frequency of outcome-specific ICD codes, counts of mentions of relevant outcome-related terms in clinical notes identified through NLP \citep{yu2018,Richesson2013}, laboratory values, or administration of certain therapies. While imperfect, well-chosen silver label surrogates that closely approximate true outcome status \citep{yu2018} enable efficient and scalable model development across large populations without requiring labor-intensive manual annotation. Throughout, we suppose that we have $q$ silver labels $\{S_j\}_{j=1}^q$ and $p$ covariates $X$. The silver labels originally proposed in PheNorm were the count of ICD codes for the outcome of interest ($S_\text{ICD}$), the count of NLP mentions of that outcome ($S_\text{NLP}$), and their sum ($S_\text{ICDNLP}$) \citep{yu2018}.

\subsection*{Prediction Methods}

\subsubsection*{PheNorm}

PheNorm \citep{yu2018} operates by assuming that the (log-transformed and normalized) silver labels each follow a mixture normal distribution with different means for outcome cases and controls. Normalization is based on the amount of healthcare utilization, often measured by clinical note count. Then the expectation-maximization (EM) algorithm \citep{dempster1977} can be used to transform silver-label predicted values into predicted outcome probabilities.

For each log-transformed and normalized silver label $S_j$ (see Supplementary Material for details), a corrupted version of $S_j$ is generated via dropout with tuning parameter $r$: $\tilde{S}_{ij} = W_{ij}S_{ij} + (1 - W_{ij})\text{mean}(S_{ij})$. Next, a linear regression model is fitted to obtain the PheNorm score $Z = (S_j, S_{-j}, X)b^*$, where 
\begin{align*}
    b^* =& \ \argmin_{b \in \mathbb{R}^{p+q}} \lVert S_j - (\tilde{S}_j, S_{-j}, X)b\rVert^2_2.
\end{align*}
Finally, predicted outcome probabilities are computed using an EM algorithm. In the E-step, the likelihood is evaluated assuming that $Z$ follows a normal distribution with parameters $(\mu_0, \sigma_0)$ among controls and $(\mu_1, \sigma_1)$ among cases, with mixture proportion $\lambda$. In the M-step, posterior probabilities are computed as $\hat{p}_S = \frac{b_1(Z)}{b_0(Z) + b_1(Z)}$, where $b_0(Z)$ and $b_1(Z)$ denote the likelihood contributions from the control and case components, respectively. An aggregate predicted probability is obtained by simple arithmetic averaging of the posterior probabilities across all silver labels, treating each label equally.

While scalable, the  accuracy of PheNorm depends on silver label quality and may decline for rare or poorly documented conditions \citep{Ferte2021}. Further, its reliance on Gaussian mixture assumptions may limit performance in rare-disease settings with sparse, zero-inflated ICD and NLP counts. 

\subsubsection*{MAP}

MAP \citep{Liao2019} extends PheNorm by incorporating multimodal ensemble learning: it fits Gaussian mixture models to both the log-transformed (normalized) silver labels and Poisson mixture models to the original count silver labels. In contrast to PheNorm, there is no dropout for the log-transformed labels. Parameters are estimated via the EM algorithm, similar to above; see the Supplementary Material for further details. 

For each silver label $S_j$, MAP produces two predicted probabilities from the EM algorithm: $\hat{p}_{S,\text{Poisson},j}$ and $\hat{p}_{S,\text{Gaussian},j}$. These can be averaged to obtain a final predicted probability $\hat{p}_{S,j}$, which can be further averaged over all silver labels to obtain the aggregate predicted probability. 
 
While MAP and PheNorm are effective for single-phenotype prediction, high-throughput applications like PheWAS require labeling hundreds to thousands of phenotypes simultaneously. This motivates the next method we consider (Figure \ref{fig:phenotyping_parallel_workflow}).

\definecolor{phenormcolor}{RGB}{180,60,100}
\definecolor{mapcolor}{RGB}{180,120,40}
\definecolor{surecolor}{RGB}{80,100,180}
\definecolor{startcolor}{RGB}{40,100,160}
\definecolor{finalcolor}{RGB}{60,140,100}

\tikzstyle{startbox} = [rectangle, rounded corners=3pt, minimum width=9cm, minimum height=1cm, 
                       text centered, draw=black, thick, fill=startcolor!20, 
                       text width=8.8cm, font=\footnotesize\bfseries]

\tikzstyle{stepbox} = [rectangle, rounded corners=2pt, minimum width=8cm, minimum height=1.5cm, 
                      text centered, draw=black, fill=white, 
                      text width=7.5cm, font=\scriptsize, inner sep=3pt]

\tikzstyle{wideStepbox} = [rectangle, rounded corners=2pt, minimum width=9cm, minimum height=2cm, 
                          text centered, draw=black, fill=white, 
                          text width=8.8cm, font=\scriptsize, inner sep=3pt]

\tikzstyle{phenormstep} = [stepbox, draw=phenormcolor, fill=phenormcolor!15]
\tikzstyle{mapstep} = [stepbox, draw=mapcolor, fill=mapcolor!15]
\tikzstyle{surestep} = [wideStepbox, draw=surecolor, fill=surecolor!15]

\tikzstyle{finalbox} = [rectangle, rounded corners=3pt, minimum width=9cm, minimum height=1.2cm, 
                       text centered, draw=black, thick, fill=finalcolor!20, 
                       text width=8.8cm, font=\footnotesize\bfseries]

\tikzstyle{methodlabel} = [rectangle, rounded corners=3pt, minimum width=4.5cm, minimum height=0.8cm,
                          text centered, font=\large\bfseries, text=white]

\tikzstyle{wideMethodlabel} = [rectangle, rounded corners=3pt, minimum width=7cm, minimum height=0.8cm,
                              text centered, font=\large\bfseries, text=white]

\tikzstyle{arrow} = [thick,->,>=stealth, line width=1.2pt]
\tikzstyle{smallarrow} = [->,>=stealth, line width=0.8pt]

\begin{figure*}[!htb]
\centering
\resizebox{0.8\textwidth}{!}{
\begin{tikzpicture}[node distance=0.6cm and 0.6cm]

\node (start) [startbox] {
    \textbf{\large Silver-Standard Labels \& Surrogate Features from EHR}\\[0.15cm]
    ICD-9/ICD-10 codes • NLP mentions • Healthcare utilization • Additional features
};

\node (phelabel) [methodlabel, fill=phenormcolor, below left=1.8cm and 0.15cm of start] {\textbf{PheNorm}};
\node (maplabel) [methodlabel, fill=mapcolor, below right=1.8cm and 0.15cm of start] {\textbf{MAP}};

\node (phe1) [phenormstep, below=0.5cm of phelabel] {
    \textbf{\large Step 1: Feature Preparation}\\[0.1cm]
    \small Extract surrogates: $S_{\text{ICD}}$, $S_{\text{NLP}}$, $S_{\text{ICDNLP}}$\\
    Normalize by healthcare utilization:\\
    $S^* = \log(1 + S) - a \log(1 + S_{\text{note}})$\\
    Optimize $a$ to minimize divergence $D(a)$
};

\node (phe2) [phenormstep, below=0.5cm of phe1] {
    \textbf{\large Step 2: Denoising via Dropout}\\[0.1cm]
    \small Create corrupted matrix with dropout rate $r$:\\
    $\widetilde{S^*}_{ij} = S^*_{ij} W_{ij} + (1-W_{ij}) \text{Mean}(S^*_j)$\\
    Fit regression: $b^* = \arg\min_b \|S^* - \widetilde{S^*}b\|_2^2$\\
    Score: $\text{PheNorm} = S^{*\top}b^*$
};

\node (phe3) [phenormstep, below=0.5cm of phe2] {
    \textbf{\large Step 3: EM for Probabilities}\\[0.1cm]
    \small E-step: $E[Y|S^*,X] = P(Y=1|S^*,X)$\\
    M-step: Update parameters\\
    Final probability:\\
    $\widehat{P}(Y=1|X) = \frac{1}{3}\sum_{\star} \text{expit}(S^{*\top}b^*_\star)$
};

\node (map1) [mapstep, below=0.5cm of maplabel] {
    \textbf{\large Step 1: Feature Extraction}\\[0.1cm]
    \small Count features:\\
    $S_{\text{ICD},i}$, $S_{\text{NLP},i}$, $S_{\text{ICDNLP},i}$\\
    Log-transform:\\
    $S_{\text{log},i} = \log(1 + S_{i})$\\
    Healthcare utilization: $S_{\text{Notelog},i}$
};

\node (map2) [mapstep, below=0.4cm of map1] {
    \textbf{\large Step 2: Mixture Models}\\[0.1cm]
    \small Poisson mixture (counts):\\
    $S_{\text{count},i}|Y_i \sim \text{Poisson}(\lambda_y)$\\
    Normal mixture (log-transformed):\\
    $S_{\text{log},i}|Y_i \sim \mathcal{N}(\mu_y, \sigma_y^2)$\\
    EM estimation for parameters
};

\node (map3) [mapstep, below=0.4cm of map2] {
    \textbf{\large Step 3: Ensemble Prediction}\\[0.1cm]
    \small Posterior probability:\\
    $P_m(Y=1|S_i) = \frac{\theta f_1(S_i)}{\theta f_1(S_i) + (1-\theta)f_0(S_i)}$\\
    Ensemble over 6 models: $P_{\text{MAP}}$\\
    Prevalence rescaling: $\hat{P}_i = g^{-1}(g(p_i) - c)$
};

\node (surelabel) [wideMethodlabel, fill=surecolor, below=2.5cm of $(phe3)!0.5!(map3)$] {\textbf{sureLDA}};

\node (sure1) [surestep, below=0.4cm of surelabel] {
    \textbf{\large Step 1: PheNorm Prior Initialization}\\[0.1cm]
    \small Run PheNorm on log-transformed features $\log(S_{k}+1)$ normalized by healthcare utilization\\
    Fit 3 Gaussian mixtures and average to get priors $\alpha_k \in [0,1]$\\
    Filter: Patients with $S_{\text{ICD},k} = 0$ get $\alpha_k = 0$ (obesity exception: both ICD and NLP = 0)
};

\node (sure2) [surestep, below=0.4cm of sure1] {
    \textbf{\large Step 2: Guided LDA Model}\\[0.1cm]
    \small Model patients as documents over $K + K_0$ phenotypes\\
    Generative process:\\
    1. Phenotype mixture: $\theta_i \sim \text{Dirichlet}(\alpha_1,\ldots,\alpha_K,1,\ldots,1)$\\
    2. Feature distribution: $\phi_k \sim \text{Dirichlet}(\beta)$ for each phenotype\\
    3. Sample features via multinomial distributions
};

\node (sure3) [surestep, below=0.4cm of sure2] {
    \textbf{\large Step 3: Feature Weighting}\\[0.1cm]
    \small Dropout regression for each phenotype $k$:\\
    $S_k^* = X_{\log}b_k + \varepsilon$\\
    Regression coefficient $b_{j,k}$ becomes weight $x_{j,k}$\\
    Negative coefficients $\rightarrow$ weight = 0\\
    (LDA cannot use negative discriminative evidence)
};

\node (sure4) [surestep, below=0.4cm of sure3] {
    \textbf{\large Step 4: Weighted Gibbs Sampling}\\[0.1cm]
    \small Weighted counts:\\
    $N_{i,k} = \sum_{j} x_{j,k} Z_{i,j,k}$ (patient-phenotype)\\
    $M_{j,k} = \sum_{i} x_{j,k} Z_{i,j,k}$ (feature-phenotype)\\
    Sample phenotype assignments via:\\
    $P_{ij;k} \propto (N_{ik}^{-ij} + \alpha_k) \cdot \frac{M_{j;k}^{-ij} + \beta}{\sum_{j'} (M_{j';k}^{-ij} + \beta)}$
};

\node (sure5) [surestep, below=0.4cm of sure4] {
    \textbf{\large Step 5: Final Probability Ensemble}\\[0.1cm]
    \small Transform LDA scores: $N_{\text{log},ik} = \log(N_{ik} + 1)$\\
    Fit Gaussian mixture adjusting for healthcare utilization\\
    Combine probabilities:\\
    $\hat{P}_{ik}^{\text{sureLDA}} = \frac{1}{4}(P^{\text{LDA}}_{ik} + P^{\text{ICD}}_{ik} + P^{\text{NLP}}_{ik} + P^{\text{ICDNLP}}_{ik})$
};

\node (final) [finalbox, below=1cm of sure5] {
    \textbf{\large Enhanced Multi-Phenotype Predictions}\\[0.15cm]
    Higher accuracy • Better feature selection • Joint phenotype modeling\\
    Robust to diverse disease characteristics • Scalable for PheWAS
};

\draw [arrow, phenormcolor] (start.south) -| (phelabel.north);
\draw [arrow, mapcolor] (start.south) -| (maplabel.north);

\draw [smallarrow, phenormcolor] (phelabel) -- (phe1);
\draw [smallarrow, phenormcolor] (phe1) -- (phe2);
\draw [smallarrow, phenormcolor] (phe2) -- (phe3);

\draw [smallarrow, mapcolor] (maplabel) -- (map1);
\draw [smallarrow, mapcolor] (map1) -- (map2);
\draw [smallarrow, mapcolor] (map2) -- (map3);

\draw [arrow, surecolor, line width=1.2pt] (phe3.south) |- (surelabel.west) 
      node[midway, below, font=\scriptsize\bfseries, text=surecolor] {PheNorm probabilities as priors};
\draw [arrow, surecolor, line width=1.2pt] (map3.south) |- (surelabel.east)
      node[midway, below, font=\scriptsize\bfseries, text=surecolor] {MAP features \& methods};

\draw [smallarrow, surecolor] (surelabel) -- (sure1);
\draw [smallarrow, surecolor] (sure1) -- (sure2);
\draw [smallarrow, surecolor] (sure2) -- (sure3);
\draw [smallarrow, surecolor] (sure3) -- (sure4);
\draw [smallarrow, surecolor] (sure4) -- (sure5);

\draw [arrow, finalcolor, line width=1.5pt] (sure5) -- (final);

\node[font=\footnotesize, text=phenormcolor, align=center, rotate=90] at ($(phe2) + (-5.5,0)$) {\textbf{Weakly-supervised denoising approach}};
\node[font=\footnotesize, text=mapcolor, align=center, rotate=90] at ($(map2) + (5.5,0)$) {\textbf{Multimodal ensemble learning}};
\node[font=\footnotesize, text=surecolor, align=center, rotate=90] at ($(sure3) + (-6.5,0)$) {\textbf{Guided probabilistic topic modeling}};

\end{tikzpicture}}
\caption{\textbf{Automated EHR phenotyping workflow with parallel processing.} Silver-standard labels from electronic health records feed into PheNorm and MAP methods in parallel. PheNorm uses weakly-supervised denoising with dropout regression and EM estimation. MAP employs multimodal ensemble learning with mixture models. sureLDA integrates both approaches: using PheNorm probabilities as Dirichlet priors and incorporating MAP features with guided LDA and weighted Gibbs sampling to produce enhanced multi-phenotype predictions for large-scale phenome-wide association studies.}
\label{fig:phenotyping_parallel_workflow}
\end{figure*}
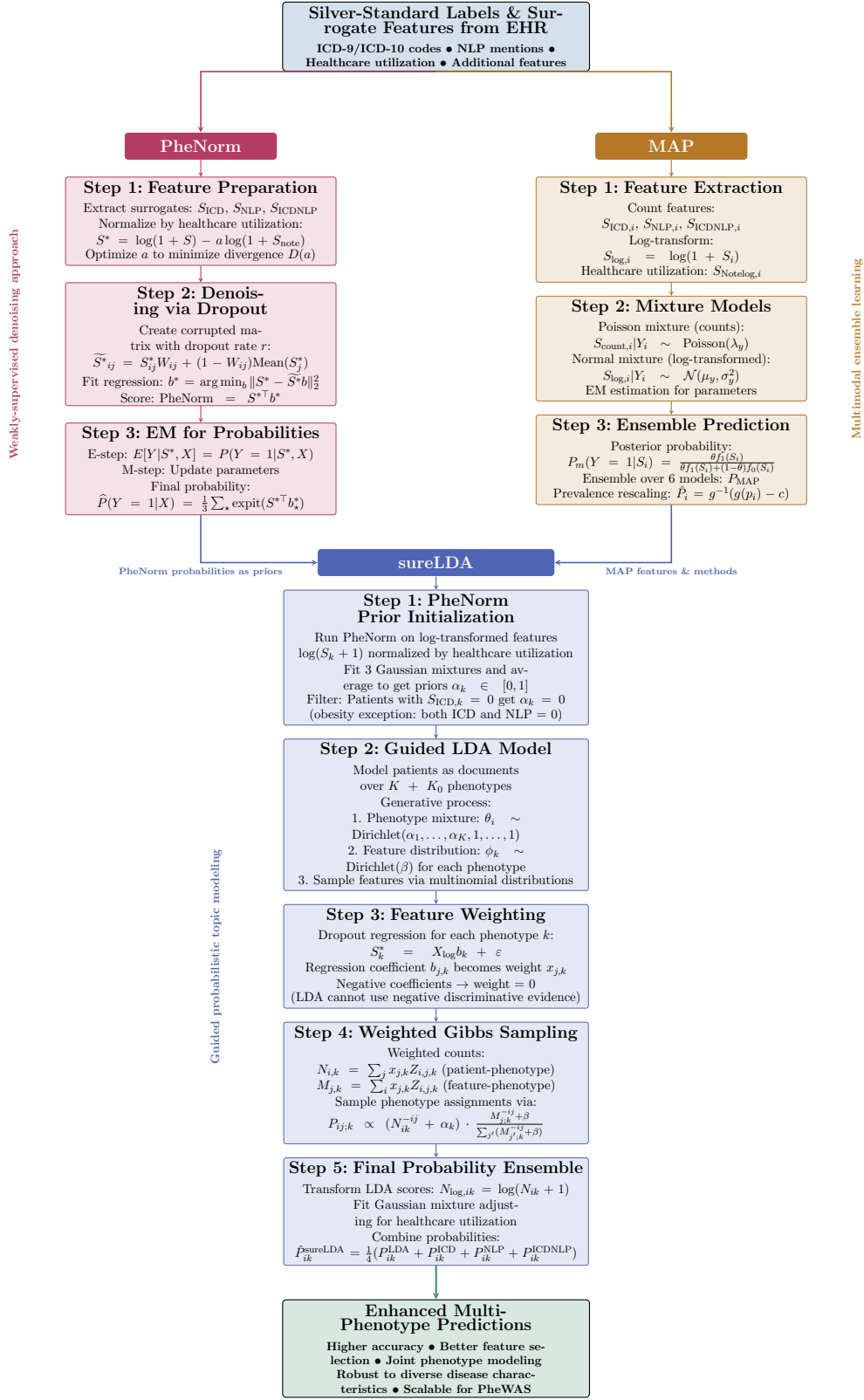

\subsubsection*{sureLDA}

SureLDA \citep{Ahuja2020} is a multi-phenotype method that jointly models K phenotypes of interest, leveraging shared information across correlated phenotypes. SureLDA can integrate phenotype probabilities from PheNorm, MAP, or another external source with latent Dirichlet allocation (LDA), modeling patients as mixtures over K latent phenotype topics.

For each phenotype, sureLDA initializes priors on log-transformed features normalized for healthcare utilization using probabilities from PheNorm, MAP, or other weakly-supervised methods. A guided LDA model is then applied with Dirichlet priors. SureLDA uses dropout regression to compute phenotype-specific feature weights, excluding negative coefficients since LDA cannot use negative discriminative evidence. Weighted Gibbs sampling estimates patient-topic and feature-topic distributions. Finally, sureLDA obtains phenotype probabilities by fitting Gaussian mixture models to the LDA scores and surrogate features, with an aggregate probability averaging across all models (see Supplementary Material for technical details). SureLDA is designed to be robust for high-dimensional phenotyping tasks and noisy surrogate data and can support simultaneous modeling of multiple phenotypes (Figure \ref{fig:phenotyping_parallel_workflow}) \citep{Ahuja2020}.

\subsection*{Simulation Methods}

We evaluated the performance of PheNorm, MAP, and sureLDA under both simple and more complex simulated EHR data conditions, varying disease prevalence and silver label informativeness to assess validity, robustness, and computational scalability across a range of scenarios.

\subsubsection*{Synthetic Dataset Architecture}

We implemented three data-generation approaches of increasing complexity, denoted as simplified, sureLDA, and complex data-generating mechanisms (Figure~\ref{fig:data_generation_flowchart}). Each incorporates essential EHR elements including patient demographics, ICD codes, and NLP features. We obtain true conditional outcome probabilities using Bayes' theorem.

Within each data-generating approach, we considered four scenarios defined by combinations of disease prevalence (rare: 5\% vs. common: 40\%) and silver-label informativeness (highly vs. weakly predictive). The rare prevalence scenario (5\%) reflects a typical prevalence for some vaccine adverse events, while the common prevalence scenario (40\%) reflects the chronic disease settings for which PheNorm, MAP, and sureLDA were originally developed. We varied the strength of association between the silver labels and true outcomes to assess whether a minimum level of predictiveness was required for adequate performance of these weakly supervised algorithms.

Full descriptions of the data-generating mechanisms are provided in Supplemental Methods S1.

\subsubsection*{Simplified Data Generation}

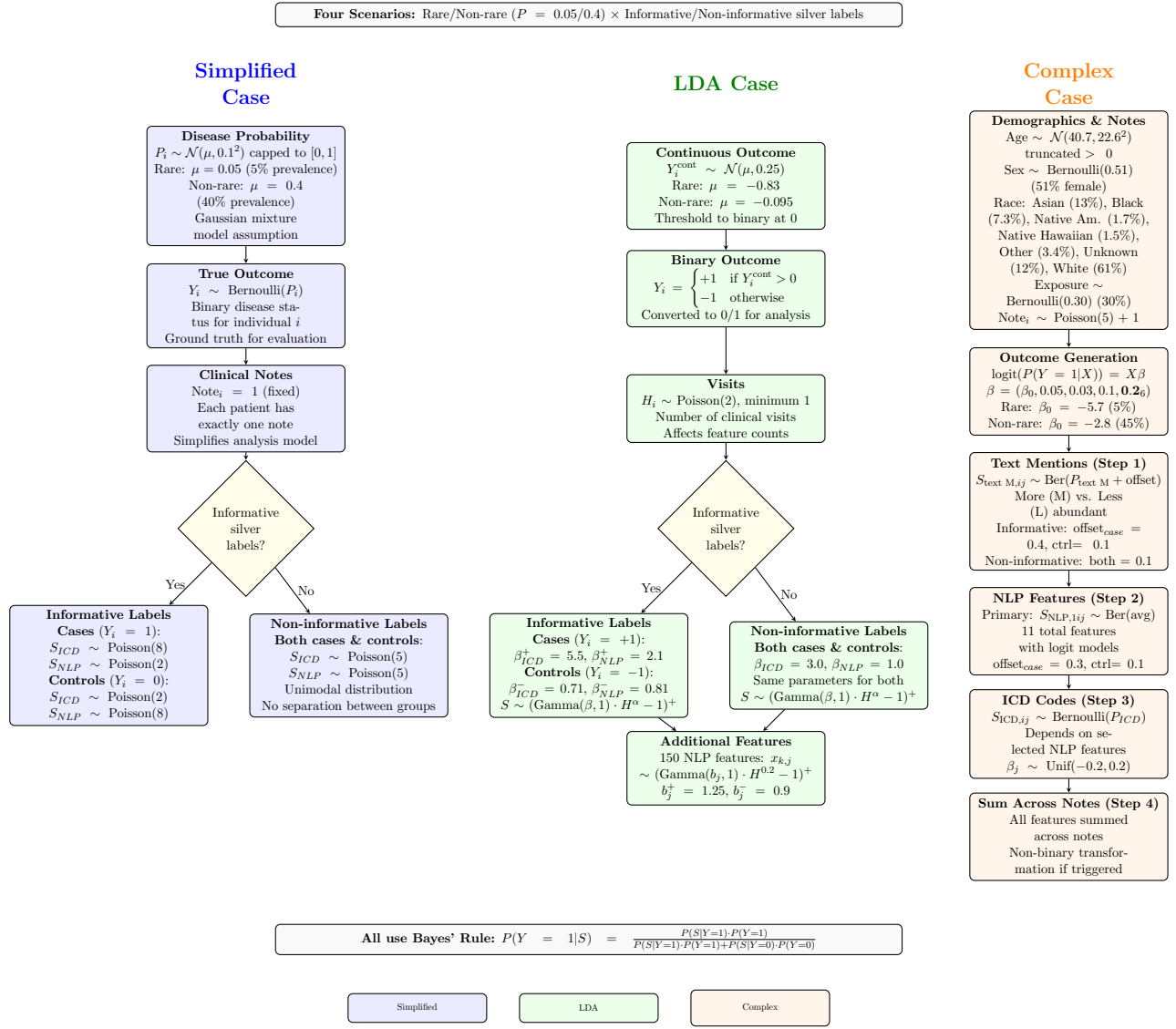
\begin{figure}[htbp]
\centering
\scalebox{0.5}{
\begin{tikzpicture}[
    node distance=5cm,
    auto,
    thick,
    block/.style={rectangle, draw, fill=blue!8, text width=5.5cm, text centered, rounded corners, minimum height=1.2cm, font=\small},
    block_lda/.style={rectangle, draw, fill=green!8, text width=5.5cm, text centered, rounded corners, minimum height=1.2cm, font=\small},
    block_complex/.style={rectangle, draw, fill=orange!8, text width=5.5cm, text centered, rounded corners, minimum height=1.5cm, font=\small},
    decision/.style={diamond, draw, fill=yellow!10, text width=2.2cm, text centered, minimum height=1cm, font=\small},
    arrow/.style={thick,->,>=stealth},
    title/.style={rectangle, draw=none, fill=none, text width=4cm, text centered, font=\Large\bfseries},
    info_box/.style={rectangle, draw, fill=gray!5, text width=18cm, text centered, rounded corners, minimum height=0.7cm, font=\small},
    phase/.style={rectangle, draw, fill=gray!12, text width=4cm, text centered, rounded corners, minimum height=0.8cm, font=\large\bfseries}
]

\node [info_box] (scenarios) at (-9, 18) {\textbf{Four Scenarios:} Rare/Non-rare ($P = 0.05/0.4$) $\times$ Informative/Non-informative silver labels};

\node [title] (title_simple) at (-19, 16) {\textcolor{blue}{Simplified Case}};
\node [title] (title_lda) at (-5, 16) {\textcolor{green!50!black}{LDA Case}};
\node [title] (title_complex) at (5, 16) {\textcolor{orange}{Complex Case}};

\node [block] (step1_simple) at (-19, 13) {\textbf{Disease Probability}\\
$P_i \sim \mathcal{N}(\mu, 0.1^2)$ capped to $[0,1]$\\
Rare: $\mu = 0.05$ (5\% prevalence)\\
Non-rare: $\mu = 0.4$ (40\% prevalence)\\
Gaussian mixture model assumption};

\node [block] (step2_simple) at (-19, 9.5) {\textbf{True Outcome}\\
$Y_i \sim \text{Bernoulli}(P_i)$\\
Binary disease status for individual $i$\\
Ground truth for evaluation};

\node [block] (step3_simple) at (-19, 6.5) {\textbf{Clinical Notes}\\
$\text{Note}_i = 1$ (fixed)\\
Each patient has exactly one note\\
Simplifies analysis model};

\node [decision] (decision_simple) at (-19, 3) {Informative\\silver\\labels?};

\node [block] (step4a_simple) at (-23, -1.0) {\textbf{Informative Labels}\\
\textbf{Cases} ($Y_i = 1$):\\
$S_{ICD} \sim \text{Poisson}(8)$\\
$S_{NLP} \sim \text{Poisson}(2)$\\
\textbf{Controls} ($Y_i = 0$):\\
$S_{ICD} \sim \text{Poisson}(2)$\\
$S_{NLP} \sim \text{Poisson}(8)$};

\node [block] (step4b_simple) at (-16, -1.0) {\textbf{Non-informative Labels}\\
\textbf{Both cases \& controls}:\\
$S_{ICD} \sim \text{Poisson}(5)$\\
$S_{NLP} \sim \text{Poisson}(5)$\\
Unimodal distribution\\
No separation between groups};

\node [block_lda] (step1_lda) at (-5, 13) {\textbf{Continuous Outcome}\\
$Y_i^{\text{cont}} \sim \mathcal{N}(\mu, 0.25)$\\
Rare: $\mu = -0.83$\\
Non-rare: $\mu = -0.095$\\
Threshold to binary at 0};

\node [block_lda] (step2_lda) at (-5, 10) {\textbf{Binary Outcome}\\
$Y_i = \begin{cases} +1 & \text{if } Y_i^{\text{cont}} > 0\\ -1 & \text{otherwise} \end{cases}$\\
Converted to 0/1 for analysis};

\node [block_lda] (step3_lda) at (-5, 6.5) {\textbf{Visits}\\
$H_i \sim \text{Poisson}(2)$, minimum 1\\
Number of clinical visits\\
Affects feature counts};

\node [decision] (decision_lda) at (-5, 3) {Informative\\silver\\labels?};

\node [block_lda] (step4a_lda) at (-9, -1.0) {\textbf{Informative Labels}\\
\textbf{Cases} ($Y_i = +1$):\\
$\beta_{ICD}^+ = 5.5$, $\beta_{NLP}^+ = 2.1$\\
\textbf{Controls} ($Y_i = -1$):\\
$\beta_{ICD}^- = 0.71$, $\beta_{NLP}^- = 0.81$\\
$S \sim (\text{Gamma}(\beta, 1) \cdot H^{\alpha} - 1)^+$};

\node [block_lda] (step4b_lda) at (-2, -1.0) {\textbf{Non-informative Labels}\\
\textbf{Both cases \& controls}:\\
$\beta_{ICD} = 3.0$, $\beta_{NLP} = 1.0$\\
Same parameters for both\\
$S \sim (\text{Gamma}(\beta, 1) \cdot H^{\alpha} - 1)^+$};

\node [block_lda] (step5_lda) at (-5, -4) {\textbf{Additional Features}\\
150 NLP features: $x_{k,j}$\\
$\sim (\text{Gamma}(b_j, 1) \cdot H^{0.2} - 1)^+$\\
$b_j^+ = 1.25$, $b_j^- = 0.9$};

\node [block_complex] (demo_complex) at (5, 12) {\textbf{Demographics \& Notes}\\
Age $\sim \mathcal{N}(40.7, 22.6^2)$ truncated $> 0$\\
Sex $\sim \text{Bernoulli}(0.51)$ (51\% female)\\
Race: Asian (13\%), Black (7.3\%), Native Am. (1.7\%),\\
Native Hawaiian (1.5\%), Other (3.4\%), Unknown (12\%), White (61\%)\\
Exposure $\sim \text{Bernoulli}(0.30)$ (30\%)\\
$\text{Note}_i \sim \text{Poisson}(5) + 1$};

\node [block_complex] (outcome_complex) at (5, 7) {\textbf{Outcome Generation}\\
$\text{logit}(P(Y=1|X)) = X\beta$\\
$\beta = (\beta_0, 0.05, 0.03, 0.1, \mathbf{0.2}_6)$\\
Rare: $\beta_0 = -5.7$ (5\%)\\
Non-rare: $\beta_0 = -2.8$ (45\%)};

\node [block_complex] (text_complex) at (5, 3.5) {\textbf{Text Mentions (Step 1)}\\
$S_{\text{text M},ij} \sim \text{Ber}(P_{\text{text M}} + \text{offset})$\\
More (M) vs. Less (L) abundant\\
Informative: offset$_{case}=0.4$, ctrl$=0.1$\\
Non-informative: both $= 0.1$};

\node [block_complex] (nlp_complex) at (5, 0) {\textbf{NLP Features (Step 2)}\\
Primary: $S_{\text{NLP},1ij} \sim \text{Ber}(\text{avg})$\\
11 total features with logit models\\
offset$_{case}=0.3$, ctrl$=0.1$};

\node [block_complex] (icd_complex) at (5, -3) {\textbf{ICD Codes (Step 3)}\\
$S_{\text{ICD},ij} \sim \text{Bernoulli}(P_{ICD})$\\
Depends on selected NLP features\\
$\beta_j \sim \text{Unif}(-0.2, 0.2)$};

\node [block_complex] (sum_complex) at (5, -6) {\textbf{Sum Across Notes (Step 4)}\\
All features summed across notes\\
Non-binary transformation if triggered};

\draw [arrow] (step1_simple) -- (step2_simple);
\draw [arrow] (step2_simple) -- (step3_simple);
\draw [arrow] (step3_simple) -- (decision_simple);
\draw [arrow] (decision_simple) -- node[left, font=\small] {Yes} (step4a_simple);
\draw [arrow] (decision_simple) -- node[right, font=\small] {No} (step4b_simple);

\draw [arrow] (step1_lda) -- (step2_lda);
\draw [arrow] (step2_lda) -- (step3_lda);
\draw [arrow] (step3_lda) -- (decision_lda);
\draw [arrow] (decision_lda) -- node[left, font=\small] {Yes} (step4a_lda);
\draw [arrow] (decision_lda) -- node[right, font=\small] {No} (step4b_lda);
\draw [arrow] (step4a_lda) -- (step5_lda);
\draw [arrow] (step4b_lda) -- (step5_lda);

\draw [arrow] (demo_complex) -- (outcome_complex);
\draw [arrow] (outcome_complex) -- (text_complex);
\draw [arrow] (text_complex) -- (nlp_complex);
\draw [arrow] (nlp_complex) -- (icd_complex);
\draw [arrow] (icd_complex) -- (sum_complex);

\node [info_box] at (-9, -9.0) {\textbf{All use Bayes' Rule:} $P(Y = 1 | S) = \frac{P(S | Y = 1) \cdot P(Y = 1)}{P(S | Y = 1) \cdot P(Y = 1) + P(S | Y = 0) \cdot P(Y = 0)}$};

\node [block, scale=0.7] at (-14, -11) {Simplified};
\node [block_lda, scale=0.7] at (-9, -11) {LDA};
\node [block_complex, scale=0.7] at (-4, -11) {Complex};

\end{tikzpicture}
}
\caption{Comprehensive flowchart illustrating simplified, LDA, and complex data generation methods for phenotyping algorithm evaluation. See Supplemental Tables S1-S3 for complete parameter specifications }
\label{fig:data_generation_flowchart}
\end{figure}

The simplified case is an idealized condition where silver labels follow simple distributions, serving as a baseline for algorithm comparison; we expect all algorithms to perform reasonably well. Individual disease probabilities follow normal distributions with prevalence-specific means, while \(S_\text{ICD}\) and \(S_\text{NLP}\) are generated from Poisson mixture distributions with parameters creating clear case-control separation in the highly predictive silver label setting or overlapping distributions in the weakly predictive silver label setting (Figure~\ref{fig:all_data_gen}). Detailed parameters are shown in Figure~\ref{fig:data_generation_flowchart}. Each patient contributes exactly one clinical note to eliminate healthcare utilization variability. Exact distributions are specified in the Supplementary Material S1.

\subsubsection*{sureLDA Data Generation}

This intermediate approach implements the same generative model used in the original sureLDA evaluation \cite{Ahuja2020}. We generate continuous latent outcomes transformed to binary disease status, with silver labels following Gamma distributions parameterized by disease status and informativeness level. Additional complexity is added with 150 auxiliary NLP variables, where each $NLP_j$ ($j = 1, \ldots, 150$) was generated from a Gamma distribution with different shape parameters for cases and for controls, scaled by healthcare utilization. These variables are not used in calculating true outcome probabilities, simulating realistic noise features in high-dimensional EHR data. This approach exhibits reduced case-control separation compared to the simplified method (Figure~\ref{fig:all_data_gen}).

\subsubsection*{Complex Data Generation}

We mimic a more realistic data-generating process in the complex approach. Patient demographics include age, sex, race, and medical history variables that influence disease probability through logistic regression. Silver label generation follows a realistic temporal sequence: first, text mentions are generated based on patient demographics and disease status; next, these mentions are processed into NLP features; finally, ICD codes are assigned based on the clinical note content. This creates dependencies characteristic of real-world phenotyping challenges. These note-level features are then summed to create patient-level silver labels  (Figure~\ref{fig:all_data_gen}).

\subsection*{Algorithm Specifications}

We evaluated multiple versions of each algorithm family (PheNorm, MAP, and sureLDA) to provide a rich set of comparisons by varying the tuning parameters; see Table~\ref{tab:algorithm_versions}.  

\begin{table}[htbp]
\centering
\small
\begin{tabularx}{\textwidth}{|l|X|}
\hline
\textbf{Method (Family-Version)} & \textbf{Description} \\
\hline
PheNorm-v1 & Standard PheNorm algorithm with Gaussian mixture model initialized with fixed variance = 1 \\
\hline
PheNorm-v2 & PheNorm with data-adaptive variance initialization (variance = half the standard deviation of normalized scores), as implemented in sureLDA\\
\hline
MAP-v1 & Standard MAP algorithm fit on all patients with non-zero ICD codes \\
\hline
MAP-v2 & MAP fit only on filter-positive patients (those meeting initial screening criteria), as implemented in sureLDA\\
\hline
sureLDA-v1 & Standard weakly-supervised sureLDA algorithm with PheNorm-v1 prior \\
\hline
sureLDA-v2 & sureLDA with MAP-v1 prior \\
\hline
sureLDA-v3 & sureLDA with PheNorm-v2 prior \\
\hline
sureLDA-v4 & sureLDA with MAP-v2 prior \\
\hline
\end{tabularx}
\caption{Phenotyping algorithm versions. v2 variants use implementation details from sureLDA: PheNorm-v2 uses data-adaptive rather than fixed variance initialization; MAP-v2 fits only on filter-positive patients rather than all patients.}
\label{tab:algorithm_versions}
\end{table}

\subsection*{Evaluation of Algorithm Performance}

For each data-generating mechanism and scenario described above, we generated 2500 independent datasets. For each dataset, to mimic a typical phenotype algorithm development study, where a large number of patient encounters are available without gold-standard labels and only a small subset has been chart-reviewed, the data were randomly split into a large training set ($n = 9800$) with labels hidden from the algorithms and a small labeled testing set ($n = 200$). Each algorithm was fit to the training data without using the true labels; then, predictions were obtained on the test data. Prediction performance metrics [area under the receiver operating characteristic curve (AUC), F1 score, precision, recall, and mean squared error between predicted and true probabilities \citep{Steyerberg2009}] were calculated on both the training and test data. We report means and standard deviations of the prediction performance metrics computed over these 2500 replications.  

To address rare instances where algorithms produced undefined predictions due to numerical instabilities in Bayesian updates (typically <5\% of individuals, occurring when true probabilities approached zero), we replaced undefined values with $\epsilon = 1\times10^{-10}$. This approach retains the full sample and avoids selection bias from excluding individuals with extreme probabilities. Results were robust to alternative approaches (Supplemental Methods~\ref{sec:handling_nas} and Figure~\ref{fig:mse_boxplot}).

\subsection*{Evaluation of Probability-Guided Chart Selection}

Our second aim was to use a vaccine safety dataset consisting of 1028 health care encounters within Kaiser Permanente Washington to evaluate whether using predicted probabilities from anaphyaxis phenotyping algorithms to guide chart selection for limited outcome validation could produce more informative validation samples compared to simple random sampling, given the same chart review budget.  

We applied each algorithm from Table~\ref{tab:algorithm_versions} to obtain predicted probabilities for anaphylaxis. We then selected a subset of encounters using a stratified approach: 80 encounters from the top 10\% of predicted probabilities, 80 from the bottom 10\%, and 40 from the middle range where algorithmic uncertainty is greatest. 

We evaluated the effectiveness of this probability-guided sampling strategy by comparing the clinical characteristics of selected encounters across different algorithms and against a random sample. We used the Kruskal--Wallis test~\cite{kruskal1952} to assess whether the distribution of structured clinical concepts captured through Concept Unique Identifiers (CUIs \citep{mccray1995}) differed significantly across sampling approaches.

\subsection*{Implementation}

Algorithms were implemented in R. The specific packages used are described in the supplemental computational environment.

All experiments were carried out on a Linux-based system equipped with an Intel Core i7 processor and 16 GB of RAM, using R version 4.1.2. Code to reproduce the simulations is available on GitHub (\url{https://github.com/Charl3456/PheNorm_public.git}).


\section*{Results}

\subsection*{Algorithm Performance in the Optimal Setting}

We first evaluated algorithm performance under optimal conditions: common outcome (40\% prevalence) and highly informative silver labels under simplified data generation. Results are in Table~\ref{tab:nonrare_informative} and the top left panel of Figures~\ref{fig:auc_performance} and \ref{fig:mse_performance}. More detail, including training-set prediction performance, is provided in Table~\ref{tab:ideal_common_info}. 

\newcolumntype{N}{>{\centering\arraybackslash}p{1.7cm}} 
\begin{table}[htbp]
\centering
\scriptsize
\begin{tabularx}{0.8\textwidth}{lNNNNN}
\toprule
\textbf{Algorithm} & \textbf{AUC} & \textbf{F1} & \textbf{Precision} & \textbf{Recall} & \textbf{Probability MSE} \\
\midrule
ICD Logit (v1)   & 0.98 (0.01) & 0.91 (0.02) & 0.91 (0.05) & 0.91 (0.04) & 0.05 (0.01) \\
PheNorm-v1     & 1.00 (<0.01) & 0.99 (0.01) & 0.98 (0.02) & 0.99 (0.01) & 0.01 (0.01) \\
PheNorm-v2     & 1.00 (<0.01) & 0.99 (0.01) & 0.98 (0.02) & 0.99 (0.01) & 0.01 (<0.01) \\
MAP-v1         & 1.00 (<0.01) & 0.98 (0.01) & 0.98 (0.02) & 0.99 (0.02) & 0.04 (0.02) \\
MAP-v2         & 1.00 (<0.01) & 0.98 (0.01) & 0.98 (0.02) & 0.98 (0.02) & 0.03 (0.02) \\
\addlinespace
sureLDA-v1     & 0.86 (0.10) & 0.82 (0.10) & 0.73 (0.16) & 0.96 (0.06) & 0.19 (0.12) \\
sureLDA-v2     & 0.73 (0.07) & 0.70 (0.07) & 0.57 (0.10) & 0.94 (0.09) & 0.35 (0.08) \\
sureLDA-v3     & 0.77 (0.04) & 0.74 (0.04) & 0.60 (0.06) & 0.96 (0.04) & 0.29 (0.05) \\
sureLDA-v4     & 0.72 (0.05) & 0.69 (0.05) & 0.55 (0.06) & 0.94 (0.10) & 0.37 (0.03) \\
\bottomrule
\end{tabularx}
\caption{Performance metrics for phenotyping algorithms in the non-rare, informative scenario with simplified data. This represents the most favorable conditions with higher disease prevalence and sufficient informative features.}
\label{tab:nonrare_informative}
\end{table}

Standard PheNorm-v1 achieved the perfect discrimination (AUC = 1), followed closely by PheNorm-v2 and both MAP variants (AUC = 1). AUCs for the four sureLDA methods were lower and varied by prior (0.72 to 0.86), with much larger SDs (0.04–0.10) than other methods ($\leq$0.01). Standard PheNorm-v1 also showed highest precision (0.98, SD = 0.02) and recall (0.99, SD = 0.01), followed by PheNorm-v2. MAP variants were also stable (v1: precision = 0.98, recall = 0.99; v2: precision = 0.98, recall = 0.98) while other sureLDA variants were highly variable (precision 0.55–0.73; recall 0.94–0.96; SDs up to 0.16). We observed similar patterns for F1 scores. Calibration metrics showed the clearest distinctions. PheNorm achieved the best calibration (MSE = 0.01, SD = 0.01),  dramatically better than all others. MSE for MAP variants was moderate (0.03 - 0.04) and poor for sureLDA variants (0.19–0.37).

\subsection*{Algorithm Performance in Intermediate Settings}

\begin{figure*}[t]
    \centering
    \includegraphics[width=\textwidth]{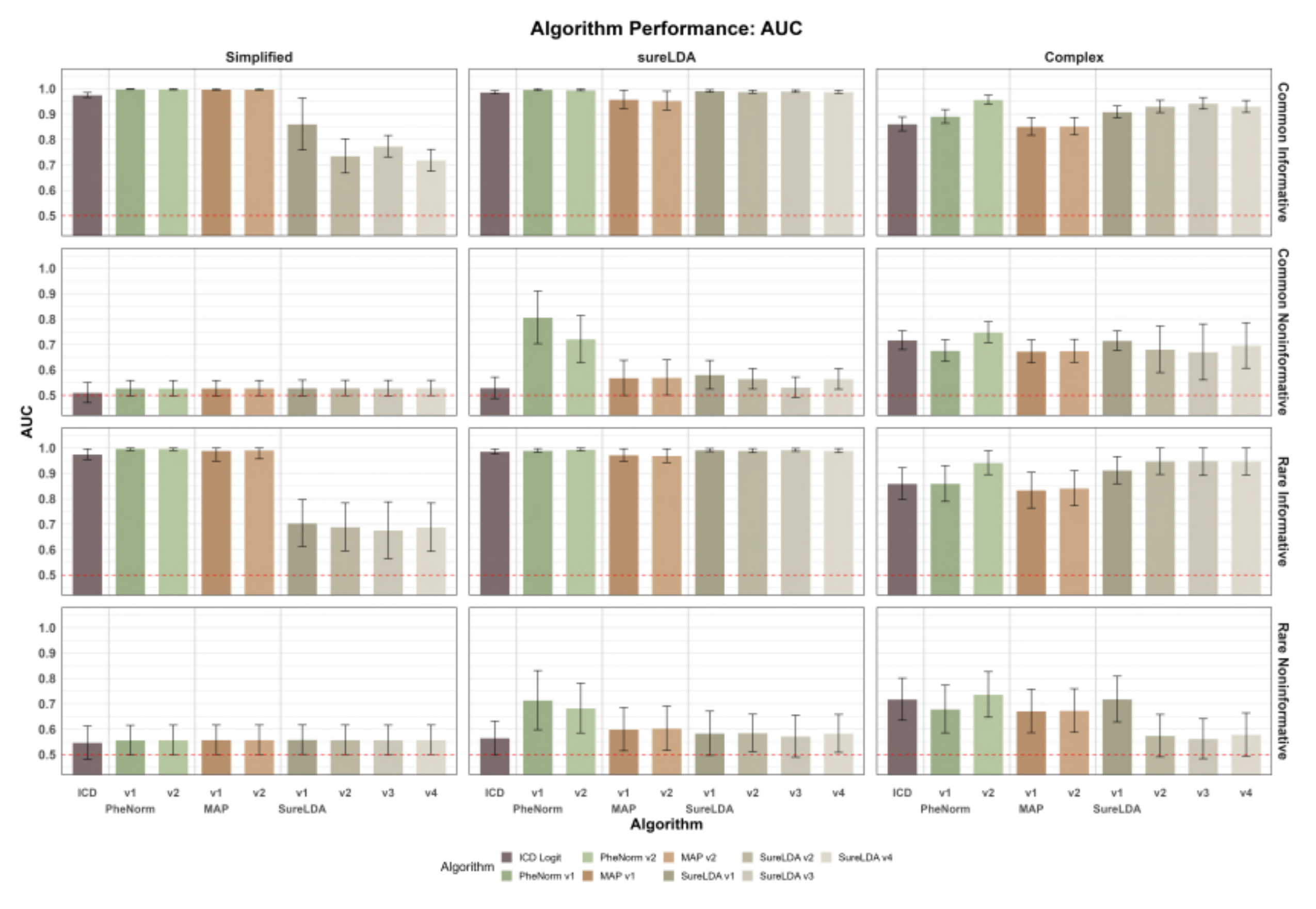}
    \caption{Algorithm performance measured by AUC across different scenarios and data types. Error bars represent one standard deviation.}
    \label{fig:auc_performance}
\end{figure*}

\begin{figure*}[t]
    \centering
    \includegraphics[width=\textwidth]{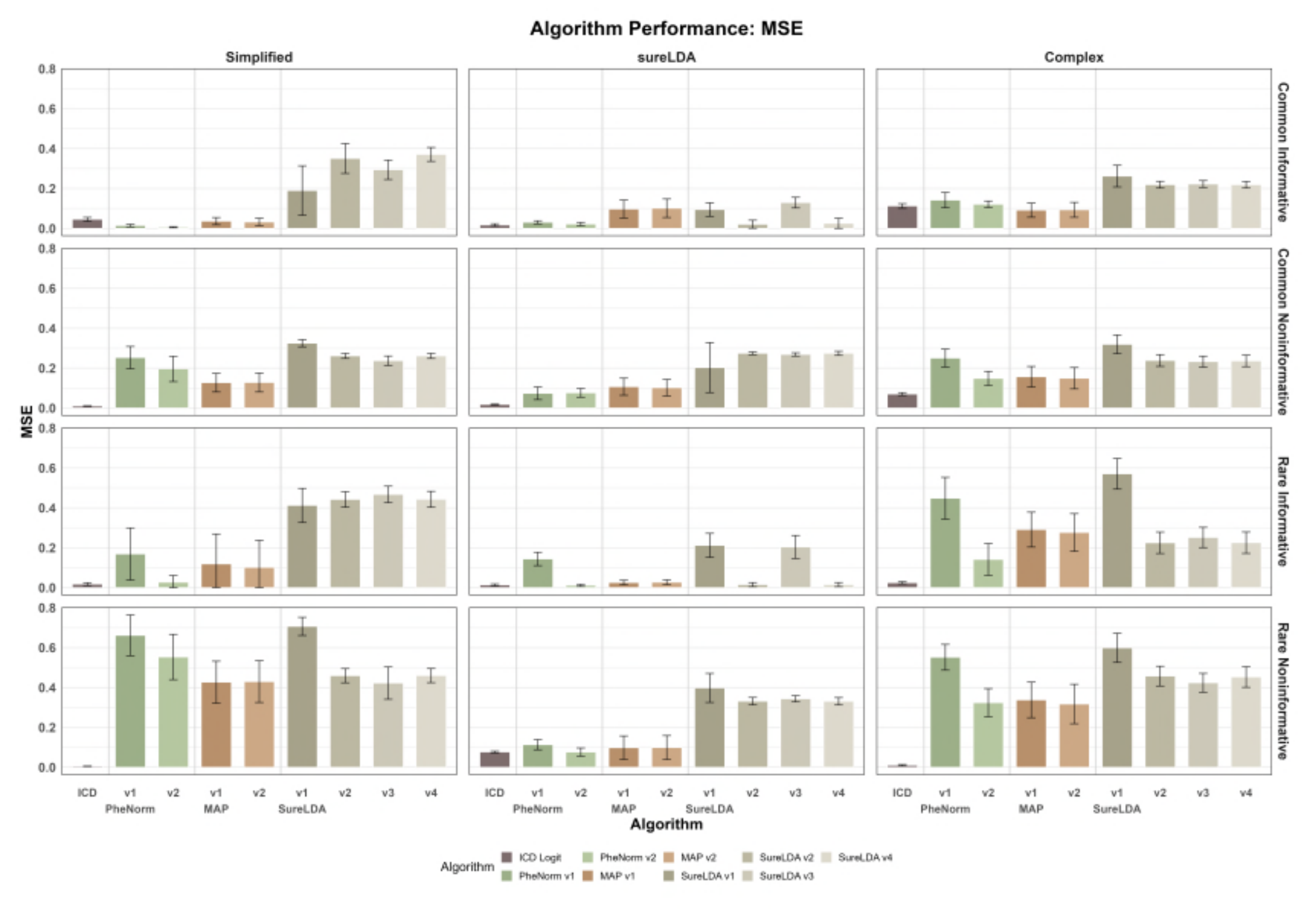}
    \caption{Algorithm performance measured by mean squared error (MSE) across different scenarios and data types. Error bars represent one standard deviation.}
    \label{fig:mse_performance}
\end{figure*}

Settings of intermediate complexity in this section include 1) the simplified data generating scenario with the three suboptimal combinations of silver label informativeness and outcome prevalence, 2) the sureLDAdata-generating scenario under all four informativeness and prevalence combinations, and 3) the complex data generating scenario except under the rare outcome and informative label combination, which we highlight in the next section as the most relevant for practical use. 
We present results for the other data-generating, outcome, and silver-label scenarios in the remaining panels of Figures~\ref{fig:auc_performance} and \ref{fig:mse_performance} and Tables~\ref{tab:rare_informative} and \ref{tab:ideal_common_noninfo}--\ref{tab:ori_rare_noninfo}.

In the simplified data scenario (first column of Figures~\ref{fig:auc_performance} and \ref{fig:mse_performance}) with non-informative silver labels and common outcome (40\% prevalence), performance substantially degraded compared with the optimal setting, with AUCs for all methods around 0.52 (Table~\ref{tab:ideal_common_noninfo}). This no better-than-chance performance was expected and confirms that all algorithms appropriately respond to the absence of meaningful signal from the silver labels. Simplified data scenario with rare outcome (5\% prevalence) and informative labels (Table~\ref{tab:ideal_rare_info}), PheNorm was remarkably stable, with near-perfect AUC and only moderate precision reduction (from 0.98 to 0.95(v1) and 0.93(v2)). MAP showed similar discrimination performance (AUC = 0.99) but larger precision decline to 0.82. SureLDA variants performed worst, with AUC dropping to 0.68--0.70 and precision to 0.14--0.16. In the rare-outcome non-informative silver label scenario of simple data generation, all algorithms performed poorly (AUC = 0.56, precision = 0.08; Table~\ref{tab:ideal_rare_noninfo}). MSE was typically larger in the non-informative versus informative silver-label scenarios, with sureLDA variants having the largest MSE.

Performance patterns differed in the sureLDA data-generation scenario (second column of Figures~\ref{fig:auc_performance} and \ref{fig:mse_performance}). Here, under the least complicated conditions (common outcome and informative labels), sureLDA variant AUCs increased to 0.99 compared with 0.73–0.86 under simplified data generation, PheNorm maintained excellence, and MAP experienced moderate degradation (Table~\ref{tab:lda_common_info}). With common outcome and non-informative labels, rather than the universal collapse to 0.52 AUC observed under simplified data generation, there was recoverable signal: PheNorm-v1 achieved AUC of 0.80, while MAP and sureLDA showed modest recovery (Table~\ref{tab:lda___common_noninformative}). With rare outcomes and informative labels, sureLDA variants again demonstrated remarkable recovery, with AUC improving to 0.99 (from 0.67-0.70 under simplified data generation for the same scenario), while PheNorm and MAP experienced modest declines from their near-perfect simplified-data performance (Table~\ref{tab:lda_rare_info}). Even under the most challenging scenario (rare outcome and non-informative labels), all methods showed improvement over simplified data generation: PheNorm-v1 achieved AUC of 0.71 (vs. 0.56 under simplified data), while MAP and sureLDA had AUCs around 0.6 (Table~\ref{tab:lda___rare_noninformative}). MSE for most methods was lower in sureLDA data-generation versus the simple data-generation scenarios, again with larger MSE for scenarios with non-informative silver-labels and rare outcomes.

Complex data generation, representing the closest to real-world conditions, revealed the most dramatic performance reorganization (third column of Figures~\ref{fig:auc_performance} and \ref{fig:mse_performance}). Under least complex conditions (40\% outcome prevalence and informative silver labels), sureLDA variants were nearly equivalent to PheNorm-v2 (AUC = 0.96), dramatically improving from simplified conditions (AUC = 0.72 - 0.86), while PheNorm-v1 degraded from 0.99 to 0.89 (Table~\ref{tab:ori_common_info}). Similar to the sureLDA data-generation scenario, there was sufficient signal in non-informative silver-label scenarios for all algorithms to achieve modest AUC (Tables~\ref{tab:ori_common_noninfo},\ref{tab:ori_rare_noninfo}). While the MSE of sureLDA improved (v1 MSE = 0.26, v2,v3, and, v4 had MSE 0.22, Table~\ref{tab:ori_common_info}), PheNorm continued to have better MSE (0.12 for v2, 0.14 for v1, Table~\ref{tab:ori_common_info}).

\subsection*{Algorithm Performance in Complex Setting Most Relevant for Practical Use}

The final scenario represents the situation that is most relevant for real-world vaccine safety outcomes: a rare disease, informative labels, and complex data generation (Tables~\ref{tab:rare_informative}, \ref{tab:ori_rare_info}). 

\newcolumntype{N}{>{\centering\arraybackslash}p{1.7cm}} 
\begin{table}[htbp]
\centering
\scriptsize
\begin{tabularx}{0.8\textwidth}{lNNNNN}
\toprule
\textbf{Algorithm} & \textbf{AUC} & \textbf{F1} & \textbf{Precision} & \textbf{Recall} & \textbf{Probability MSE} \\
\midrule
ICD Logit (v1)   & 0.86 (0.06) & 0.34 (0.12) & 0.23 (0.13) & 0.81 (0.14) & 0.03 (0.01) \\
PheNorm-v1     & 0.86 (0.07) & 0.41 (0.16) & 0.30 (0.17) & 0.82 (0.13) & 0.45 (0.10) \\
PheNorm-v2     & 0.94 (0.05) & 0.58 (0.19) & 0.46 (0.22) & 0.92 (0.09) & 0.14 (0.08) \\
MAP-v1         & 0.84 (0.07) & 0.31 (0.12) & 0.20 (0.11) & 0.88 (0.11) & 0.29 (0.09) \\
MAP-v2         & 0.84 (0.07) & 0.33 (0.12) & 0.21 (0.11) & 0.87 (0.11) & 0.28 (0.10) \\
\addlinespace
sureLDA-v1     & 0.91 (0.05) & 0.47 (0.16) & 0.34 (0.17) & 0.89 (0.10) & 0.57 (0.08) \\
sureLDA-v2     & 0.95 (0.05) & 0.62 (0.20) & 0.51 (0.23) & 0.92 (0.09) & 0.23 (0.05) \\
sureLDA-v3     & 0.95 (0.05) & 0.64 (0.20) & 0.53 (0.25) & 0.94 (0.08) & 0.25 (0.05) \\
sureLDA-v4     & 0.95 (0.05) & 0.62 (0.20) & 0.51 (0.24) & 0.92 (0.09) & 0.23 (0.05) \\
\bottomrule
\end{tabularx}
\caption{Performance metrics for phenotyping algorithms in a rare, informative scenario. This represents a real-world, complex case where these algorithms might be deployed.} 
\label{tab:rare_informative}
\end{table}

Under these conditions, sureLDA variants emerged as the strongest performers for classification, with sureLDA-v3 achieving the highest discrimination (AUC = 0.95, F1 = 0.64, precision = 0.53), followed closely by sureLDA-v2 (AUC = 0.95, F1 = 0.62, precision = 0.51). These substantial performance gains under complex conditions (+0.28 AUC for v3, +0.26 AUC for v2 compared to simplified data generation; Table~\ref{tab:ideal_rare_info}) demonstrate an ability to leverage realistic data complexity even under severe class imbalance. However, PheNorm-v2 achieved the lowest MSE with an AUC nearly as good as sureLDA-v2 (0.94). Both MAP variants had low AUC and large MSE in this setting.

\subsection*{Generalization from Training to Test Data}

Algorithm generalization patterns varied across data-generating mechanisms. Under simplified conditions, PheNorm and MAP generalized nearly perfectly (train-test AUC differences <0.01), while sureLDA variants showed substantial overfitting (e.g., sureLDA-v3: training AUC=1.00 vs. test AUC=0.77 in optimal scenario; sureLDA-v2: 0.99 vs. 0.69 in rare scenario). However, as data complexity increased, sureLDA's generalization improved markedly. Under complex data generation with rare outcomes and informative labels, all algorithms showed excellent generalization with train-test AUC differences consistently <0.03. These patterns reveal that sureLDA's architectural complexity, while prone to overfitting under simplified conditions, enables learning of generalizable patterns when data structure matches real-world EHR complexity. Conversely, PheNorm and MAP's simpler architectures provide consistent generalization across all conditions.

\subsection*{Probability-Guided Chart Selection}
We present results for PheNorm-v1, MAP-v1, sureLDA-v1, and sureLDA-v3 in Table \ref{tab:phenotyping_comparison}; the remaining algorithms are in Table~\ref{tab:phenotyping_comparison_all}. For comparison, we also had a simple random sample of 145 encounters that had been previously manually reviewed to serve as a reference. 

Compared to the simple random sample, algorithm-guided samples maintained reasonable demographic representativeness but differed with respect to NLP variables. Racial/ethnic composition was predominantly White (63-73\%) followed by Asian (9-14\%) and African American (5-9\%) patients. Algorithm-guided samples consistently showed higher frequencies of anaphylaxis mentions (4.8-6.6 vs. 3.12 per patient), antihistamine administration (71-76.5\% vs. 62.8\%), and epinephrine administration (68.5-74\% vs. 62.1\%) compared to the simple random sample. Analysis of 95 medical concepts (CUIs) extracted from clinical notes showed 21 with statistically significant differences across sampling approaches (Kruskal-Wallis test p-value < 0.05, effect sizes up to 3.4), with algorithm-guided samples showing higher concentrations of concepts related to allergic reactions, emergency treatments, and symptom presentations.

\begin{table*}[htbp]
\centering
\caption{Comparison of Patient Characteristics Across Different Phenotyping Methods}
\label{tab:phenotyping_comparison}
\resizebox{\textwidth}{!}{
\begin{tabular}{lccccc}
\toprule
\textbf{Characteristic} & \textbf{Simple Random Sample} & \textbf{MAP-v1} & \textbf{PheNorm-v1} & \textbf{sureLDA-v1} & \textbf{sureLDA-v3}\\
\midrule
\cellcolor{gray!10}Number of Patients & \cellcolor{gray!10}145 & \cellcolor{gray!10}200 & \cellcolor{gray!10}200 & \cellcolor{gray!10}200 & \cellcolor{gray!10}200 \\
Age in Years [Mean (SD)] & 49.1 (19) & 48.3 (17.1) & 50.3 (18.1) & 49.1 (17.7) & 42.6 (18.1) \\
\cellcolor{gray!10}Sex (Female) & \cellcolor{gray!10}95 (65.5\%) & \cellcolor{gray!10}140 (70\%) & \cellcolor{gray!10}140 (70\%) & \cellcolor{gray!10}151 (75.5\%) & \cellcolor{gray!10}134 (67\%) \\
Race: African American & 7 (4.8\%) & 17 (8.5\%) & 16 (8\%) & 11 (5.5\%) & 15 (7.5\%) \\
\cellcolor{gray!10}Race: Asian & \cellcolor{gray!10}18 (12.4\%) & \cellcolor{gray!10}20 (10\%) & \cellcolor{gray!10}18 (9\%) & \cellcolor{gray!10}18 (9\%) & \cellcolor{gray!10}27 (13.5\%) \\
Race: Native Hawaiian or Pacific Islander & 3 (2.1\%) & 4 (2\%) & 5 (2.5\%) & 3 (1.5\%) & 3 (1.5\%) \\
\cellcolor{gray!10}Race: Alaska Native or Native American & \cellcolor{gray!10}1 (0.7\%) & \cellcolor{gray!10}2 (1\%) & \cellcolor{gray!10}2 (1\%) & \cellcolor{gray!10}0 (0\%) & \cellcolor{gray!10}1 (0.5\%) \\
Race: Other & 5 (3.4\%) & 8 (4\%) & 9 (4.5\%) & 7 (3.5\%) & 9 (4.5\%) \\
\cellcolor{gray!10}Race: Unknown & \cellcolor{gray!10}19 (13.1\%) & \cellcolor{gray!10}20 (10\%) & \cellcolor{gray!10}21 (10.5\%) & \cellcolor{gray!10}16 (8\%) & \cellcolor{gray!10}18 (9\%) \\
Race: White & 92 (63.4\%) & 129 (64.5\%) & 129 (64.5\%) & 145 (72.5\%) & 127 (63.5\%) \\
\addlinespace
\cellcolor{gray!10}Received Antihistamine & \cellcolor{gray!10}91 (62.8\%) & \cellcolor{gray!10}147 (73.5\%) & \cellcolor{gray!10}153 (76.5\%) & \cellcolor{gray!10}148 (74\%) & \cellcolor{gray!10}142 (71\%) \\
Received Epinephrine & 90 (62.1\%) & 143 (71.5\%) & 137 (68.5\%) & 139 (69.5\%) & 148 (74\%) \\
\addlinespace
\cellcolor{gray!10}Anaphylaxis ICD Codes [Mean (SD)] & \cellcolor{gray!10}1.14 ± 0.91 & \cellcolor{gray!10}1.05 ± 1.25 & \cellcolor{gray!10}0.96 ± 1.18 & \cellcolor{gray!10}0.88 ± 1.04 & \cellcolor{gray!10}0.86 ± 1.03 \\
Anaphylaxis Mentions in Notes [Mean (SD)] & 3.12 ± 4.47 & 4.8 ± 10.31 & 5.24 ± 9.47 & 6.59 ± 10.04 & 5.4 ± 9.87 \\
\cellcolor{gray!10}Clinical Concepts (CUIs) in Notes [Mean (SD)] & \cellcolor{gray!10}1.37 ± 1.58 & \cellcolor{gray!10}1.5 ± 2.35 & \cellcolor{gray!10}1.94 ± 2.38 & \cellcolor{gray!10}2.32 ± 2.58 & \cellcolor{gray!10}1.85 ± 2.36 \\
\bottomrule
\end{tabular}
}
\end{table*}

\subsection*{Data Availability}
The dataset analyzed during this study is not publicly available because it contains detailed information from the electronic health record and is governed by HIPAA. Data are however available from the authors upon reasonable request, with permission of the health system and fully executed data use agreement.


\section*{Discussion}

In this paper, we present a comprehensive evaluation of three widely used phenotyping algorithms (PheNorm, MAP, and sureLDA) across EHR-based data generation conditions relevant to vaccine safety surveillance. We demonstrate that algorithm performance is highly context-dependent, with no single method maintaining superiority across all scenarios considered here. 

Our results highlight the importance of evaluating methods in contexts similar to those where they will be deployed. Under optimal conditions with simplified data, high prevalence, and reliable labels, PheNorm and MAP achieved near-perfect performance while sureLDA showed unexpected instability despite the favorable setting. This establishes important baselines, but these idealized conditions rarely reflect real-world EHR environments. As data complexity increased, the hierarchies of performance changed. SureLDA variants, which struggled under simplified conditions, demonstrated  robustness to data complexity, with classification performance increasing to an AUC of 0.95 in the rare-outcome complex-data setting. Conversely, PheNorm and MAP, while maintaining reasonable performance, showed either modest gains or decreases in performance.

The trade-off that we observed between discrimination and calibration quality highlights that deployment decisions must align with downstream analytical goals\citep{Pivovarov2015}. While PheNorm was best calibrated under controlled conditions, MAP offered more consistent calibration across scenarios. SureLDA didn't achieved the best calibration among all methods. Vaccine safety surveillance prioritizing case identification may favor methods optimizing discrimination, whereas epidemiologic studies using predicted probabilities in regression models require careful calibration assessment regardless of classification performance.

These findings have direct implications for using phenotyping algorithms in vaccine safety surveillance. For standard phenotyping tasks under complex conditions with reliable labels, PheNorm-v2 with data-adaptive variance initialization achieves optimal performance (AUC = 0.94, precision = 0.46, calibration = 0.14), while the sureLDA variant using PheNorm-v2 priors (sureLDA-v3) provides a competitive alternative (AUC = 0.95, precision = 0.64, calibration 0.25). Algorithm performance varied substantially based on configuration choices, with AUC differences up to 0.28 points depending on tuning parameters and scenario complexity, highlighting that analysts should evaluate multiple configurations rather than relying on default settings. 

To complement our simulation study, we conducted a proof-of-concept analysis using real-world data from 1,028 patient encounters with potential anaphylaxis. Algorithm-guided sampling consistently identified cases with more detailed clinical documentation and higher rates of appropriate emergency treatments compared to random sampling, suggesting predicted probabilities offer practical advantages for validation studies by efficiently identifying cases with comprehensive clinical information.

There are limitations to our approach. First, while we defined a broad set of scenarios to evaluate algorithm performance, our simulations cannot fully capture real-world clinical data heterogeneity. Our proof-of-concept analysis demonstrated practical feasibility of probability-guided sampling but was limited to a single health system (Kaiser Permanente Washington) and one outcome (anaphylaxis), potentially limiting generalizability. We lacked gold-standard chart review outcomes for probability-guided samples, and the 145 manually reviewed charts represent a separate random sample rather than comprehensive validation of the algorithm-guided approach. Finally, we evaluated only three weakly-supervised methods and their variants; other phenotyping approaches may perform differently under rare-outcome conditions.

Several research directions emerge from this work. First, validation using real EHR datasets with adjudicated outcomes across multiple health systems would strengthen evidence for operational performance. Second, developing adaptive methods that automatically adjust to varying silver label quality, outcome prevalence, and data complexity would enhance practical utility. Third, prospective validation with probability-guided chart review would enable direct assessment of whether this sampling strategy improves validation efficiency over random sampling. Finally, extending this framework to additional rare adverse events would establish broader applicability and identify condition-specific performance patterns to guide method selection in vaccine safety surveillance.


\section*{Conclusion}
This study underscores the need for nuanced algorithm evaluation tailored to public health surveillance demands. Organizations planning phenotyping algorithm deployment should prioritize evaluation under data conditions reflecting real-world EHR complexity rather than relying on simplified conditions alone. Algorithm choice and configuration should reflect not only theoretical performance but empirical robustness under operational stress, including: class imbalance, variable silver label quality, and realistic data dependencies. The optimal algorithm depends on downstream use, because there can be a trade-off between discrimination and calibration. With careful selection, appropriate tuning, and thorough validation, phenotyping systems can support timely, accurate detection of vaccine-related adverse events and inform public health decision-making.


\section*{Funding}
This project was supported by Task Order 75D30122F00001 from the U.S. Centers for Disease Control and Prevention (CDC). The content is solely the responsibility of the authors and does not necessarily represent the official views of CDC or the U.S. Government.

\renewcommand{\thefigure}{S\arabic{figure}}
\renewcommand{\thetable}{S\arabic{table}}
\renewcommand{\thesection}{S\arabic{section}}
\setcounter{figure}{0}
\setcounter{table}{0}
\setcounter{section}{0}
\section{Supplemental Methods}

\subsection{Method Technical Details}
\subsubsection{PheNorm Technical Details}

The log-linear normalization for silver label $S_j$ is defined as
\[
S_{\log,j} = \log(1 + S_j) - a \log(1 + S_{\text{note}}),
\]
where $S_{\text{note}}$ denotes the number of clinical notes. The parameter $a$ is selected to minimize the divergence
\[
D(a) = \int \left| F_{\text{emp}}(S_{\log}; a) - U(z; k, \ell_1, \sigma, \ell_0) \right| \, dS_{\log},
\]
where $F_{\text{emp}}$ is the empirical distribution of $S_{\log}$ and $U(\cdot)$ denotes a reference distribution.

A corrupted version of the normalized silver label is generated via dropout with tuning parameter $r$:
\[
\widetilde{S}_{\log,ij}
= S_{\log,ij} W_{ij} + (1 - W_{ij}) \, \text{Mean}(S_{\log,j}),
\]
where $W_{ij} \sim \text{Bernoulli}(1 - r)$.

Regression coefficients are estimated by solving
\[
b^* = \arg \min_b \, \left\| S_{\log,j} - (\widetilde{S}_{\log}, S_{\log,-j}, X)b \right\|_2^2,
\]
where $S_{\log,-j}$ denotes the remaining silver labels and $X$ represents additional covariates. The resulting \textsc{PheNorm} score is then computed as
\[
Z_j = (S_{\log,j}^\top, S_{\log,-j}, X)b^*.
\]

Predicted outcome probabilities are obtained using an EM algorithm for a two-component normal mixture model. In the E-step, the likelihood for a given score $z$ is
\begin{align*}
\ell(z) = b_0(z) + b_1(z)
&= \frac{\lambda}{\sqrt{\sigma_0^2}}
\exp\left\{ -\frac{(z - \mu_0)^2}{2\sigma_0^2} \right\} \\
&\quad + \frac{1 - \lambda}{\sqrt{\sigma_1^2}}
\exp\left\{ -\frac{(z - \mu_1)^2}{2\sigma_1^2} \right\},
\end{align*}
where $(\mu_0, \sigma_0)$ parameterize the control distribution, $(\mu_1, \sigma_1)$ parameterize the case distribution, and $\lambda$ is the mixture proportion.

For multiple silver labels, such as $S_{\text{ICD}}$, $S_{\text{NLP}}$, and $S_{\text{ICDNLP}}$, the final predicted probability is computed as
\[
\widehat{P}(Y = 1 \mid X)
= \frac{1}{3}
\sum_{S \in \{S_{\text{ICD}}, S_{\text{NLP}}, S_{\text{ICDNLP}}\}}
\hat{p}_S.
\]
\subsubsection{MAP Technical Details}

The \textsc{MAP} method extracts count-based silver-label features for each subject $i$, including
$S_{\text{ICD},i}$, $S_{\text{NLP},i}$, and $S_{\text{ICDNLP},i}$. A log transformation is applied to each feature,
\[
S_{\log,i} = \log(1 + S_i),
\]
and a log-transformed healthcare utilization variable $S_{\text{Note}\log,i}$ is additionally included.

For each feature, \textsc{MAP} fits mixture models to characterize the distributions among cases and controls. A Poisson mixture model is used for count-based features,
\[
S_{\text{count},i} \mid Y_i \sim \text{Poisson}(\lambda_y),
\]
and a Gaussian mixture model is used for log-transformed features,
\[
S_{\log,i} \mid Y_i \sim \mathcal{N}(\mu_y, \sigma_y^2),
\]
where $y \in \{0,1\}$ denotes disease status.

Posterior probabilities are computed as
\[
P_m(Y = 1 \mid S_i)
= \frac{\theta f_1(S_i)}{\theta f_1(S_i) + (1 - \theta) f_0(S_i)},
\]
where $f_1(S_i)$ and $f_0(S_i)$ denote the conditional densities under $Y_i = 1$ and $Y_i = 0$, respectively, and $\theta = P(Y = 1)$ is the disease prevalence.

When multiple models are fitted (e.g., Poisson and Gaussian mixtures applied to $S_{\text{ICD}}$, $S_{\text{NLP}}$, and $S_{\text{ICDNLP}}$), the ensemble posterior probability is computed by averaging across six models:
\[
P_{\text{MAP}}(Y = 1 \mid S_i)
= \frac{1}{6} \sum_{m=1}^6 P_m(Y = 1 \mid S_i).
\]

Finally, prevalence adjustment is applied to calibrate the ensemble posterior probabilities. Let $p_i = P_{\text{MAP}}(Y = 1 \mid S_i)$ denote the ensemble-averaged posterior probability. The adjusted probability is defined as
\[
\hat{P}_i = g^{-1}(g(p_i) - c),
\]
where $g(p) = \log\!\left(\frac{p}{1-p}\right)$ is the logit transformation and $c$ is chosen such that
\[
\frac{1}{n} \sum_{i=1}^n g^{-1}(g(p_i) - c) = \hat{\theta},
\]
with $\hat{\theta}$ denoting the target prevalence.

\subsubsection{sureLDA Technical Details}

For each phenotype $k \in \{1, \ldots, K\}$, \textsc{sureLDA} initializes priors using log-transformed silver-label features,
\[
S_{\log,k} = \log(S_k + 1),
\]
where silver labels may include any proxy measures available in the EHR, such as ICD codes, NLP mentions, laboratory values, or medication records.

A guided LDA model is then fit. Phenotype-topic proportions are drawn from a Dirichlet prior,
\[
\theta \sim \text{Dirichlet}(\alpha_1, \ldots, \alpha_K, 1, \ldots, 1),
\]
where $\alpha_k$ denotes the prior weight for phenotype $k$. Feature-topic distributions are drawn as
\[
\phi_k \sim \text{Dirichlet}(\beta),
\]
with $\beta = 1$ representing a non-informative prior over all features. Feature assignments are sampled from multinomial distributions.

To estimate feature weights, a dropout regression is performed for each phenotype $k$,
\[
S_{\log,k} = X_{\log} b_k + \varepsilon,
\]
where $b_{j,k}$ denotes the regression coefficient for feature $j$. Feature weights are defined as $x_{j,k} = b_{j,k}$, with negative coefficients truncated to zero.

Weighted Gibbs sampling is then conducted using the weighted counts
\[
N_{i,k} = \sum_j x_{j,k} Z_{i,j,k},
\qquad
M_{j,k} = \sum_i x_{j,k} Z_{i,j,k},
\]
where $N_{i,k}$ is the weighted count of phenotype $k$ for patient $i$, $M_{j,k}$ is the weighted count of feature $j$ for phenotype $k$, and $Z_{i,j,k}$ is the latent indicator denoting assignment of feature $j$ in patient $i$ to phenotype $k$.

Phenotype assignments are sampled according to
\[
P_{i,j;k} \propto
\left(N_{i,k}^{-ij} + \alpha_k \right)
\cdot
\frac{M_{j,k}^{-ij} + \beta}
{\sum_{j'} \left(M_{j',k}^{-ij} + \beta \right)},
\]
where the superscript $-ij$ denotes exclusion of the current feature-patient assignment.

For final probability estimation, LDA-derived scores are log-transformed as
\[
N_{\log,ik} = \log(N_{ik} + 1),
\]
and a two-component Gaussian mixture model is fit to $N_{\log,ik}$ while adjusting for log-transformed healthcare utilization,
\[
H_{\log,i} = \log(H_i),
\]
yielding posterior probabilities $P^{\text{LDA}}_{ik}$. For each of $M$ silver-standard surrogate features, normalized surrogates $S_{\log,m,ik}$ are similarly modeled using Gaussian mixture models adjusted for $H_{\log,i}$, producing posterior probabilities $P^{(m)}_{ik}$.

The final \textsc{sureLDA} probability is computed as
\[
\widehat{P}^{\text{sureLDA}}_{ik}
=
\frac{1}{M + 1}
\left(
P^{\text{LDA}}_{ik}
+
\sum_{m=1}^{M} P^{(m)}_{ik}
\right).
\]

\subsection{Data Generation Methods}

We implemented three data-generating mechanisms of increasing complexity to evaluate phenotyping algorithm performance under different scenarios. Each method generates data for $n=10,000$ individuals with a test set of 200 individuals, across four scenarios defined by disease prevalence (rare: 5\% vs. common: 40\%) and silver-label informativeness (highly vs. weakly predictive).

\subsubsection{Method 1: Simplified Data Generation}

\paragraph{Overview}
This idealized case uses direct probability generation and simple Poisson count generation to establish baseline algorithm performance. Our goal was to follow the distributional assumptions of PheNorm as closely as possible.

\paragraph{Parameters:}
\begin{itemize}[leftmargin=*]
    \item $n = 10,000$ individuals
    \item Test set size = 200
    \item Scenarios: rare\_info, rare\_noninfo, nonrare\_info, nonrare\_noninfo
\end{itemize}

\paragraph{Step 1: Generate initial probability}
\begin{itemize}[leftmargin=*]
    \item \textbf{Rare scenarios}:
    \begin{itemize}[leftmargin=2em]
        \item $q_j \sim \mathcal{N}(0.05, 0.1^2)$, truncated to $[0, 1]$
    \end{itemize}
    
    \item \textbf{Non-rare scenarios}:
    \begin{itemize}[leftmargin=2em]
        \item $q_j \sim \mathcal{N}(0.4, 0.1^2)$, truncated to $[0, 1]$
    \end{itemize}
\end{itemize}

\paragraph{Step 2: Generate true outcome}
\begin{itemize}[leftmargin=*]
    \item $Y_j \sim \text{Bernoulli}(\text{initial\_probability}_j)$
    \item $S_{\text{note},j} \sim 1 + \mathcal{N}(0, 0.001^2)$ (noisy note variable for MAP)
\end{itemize}

\paragraph{Step 3: Generate silver label counts}
\begin{itemize}[leftmargin=*]
    \item \textbf{Informative scenarios (bimodal counts)}:
    \begin{itemize}[leftmargin=2em]
        \item If $Y_j = 1$ (case):
        \begin{itemize}[leftmargin=2em]
            \item $S_{\text{ICD},j} \sim \text{Poisson}(8)$
            \item $S_{\text{NLP},j} \sim \text{Poisson}(8 \times 1.5) = \text{Poisson}(12)$
        \end{itemize}
        \item If $Y_j = 0$ (control):
        \begin{itemize}[leftmargin=2em]
            \item $S_{\text{ICD},j} \sim \text{Poisson}(2)$
            \item $S_{\text{NLP},j} \sim \text{Poisson}(2 \times 1.5) = \text{Poisson}(3)$
        \end{itemize}
    \end{itemize}
    
    \item \textbf{Non-informative scenarios (unimodal counts)}:
    \begin{itemize}[leftmargin=2em]
        \item If $Y_j = 1$ (case):
        \begin{itemize}[leftmargin=2em]
            \item $S_{\text{ICD},j} \sim \text{Poisson}(5)$
            \item $S_{\text{NLP},j} \sim \text{Poisson}(5 \times 1.5) = \text{Poisson}(7.5)$
        \end{itemize}
        \item If $Y_j = 0$ (control):
        \begin{itemize}[leftmargin=2em]
            \item $S_{\text{ICD},j} \sim \text{Poisson}(5)$
            \item $S_{\text{NLP},j} \sim \text{Poisson}(5 \times 1.5) = \text{Poisson}(7.5)$
        \end{itemize}
    \end{itemize}
\end{itemize}

\paragraph{Step 4: Update initial probability to get true probability}
\begin{itemize}[leftmargin=*]
    \item Calculate case likelihood:
    \begin{itemize}[leftmargin=2em]
        \item $p_{\text{case},j} = q_j \times P(S_{\text{ICD},j} | \text{case}) \times P(S_{\text{NLP},j} | \text{case})$
        \item Where $P(S_{\text{ICD},j} | \text{case})$ = Poisson density at observed ICD count with $\lambda_{\text{case}}$
        \item Where $P(S_{\text{NLP},j} | \text{case})$ = Poisson density at observed NLP count with $\lambda_{\text{case}} \times 1.5$
    \end{itemize}
    
    \item Calculate control likelihood:
    \begin{itemize}[leftmargin=2em]
        \item $p_{\text{control},j} = (1 - q_j) \times P(S_{\text{ICD},j} | \text{control}) \times P(S_{\text{NLP},j} | \text{control})$
    \end{itemize}
    
    \item Final probability:
    \begin{itemize}[leftmargin=2em]
        \item If $p_{\text{case},j} + p_{\text{control},j} = 0$: $p_j$ = $q_j$
        \item Otherwise: $p_j = \frac{p_{\text{case},j}}{p_{\text{case},j} + p_{\text{control},j}}$
    \end{itemize}
\end{itemize}

\subsubsection{Method 2: LDA-Inspired Data Generation}

\paragraph{Overview}
This intermediate approach implements latent variable structures using gamma distributions for counts, matching the generative model used in sureLDA evaluation \citep{Ahuja2020}.

\paragraph{Parameters:}
\begin{itemize}[leftmargin=*]
    \item $n = 10,000$ individuals
    \item Test set size = 200
    \item Scenarios: rare\_info, rare\_noninfo, nonrare\_info, nonrare\_noninfo
\end{itemize}

\paragraph{Step 1: Generate continuous and binary outcomes}
\begin{itemize}[leftmargin=*]
    \item \textbf{Rare scenarios}:
    \begin{itemize}[leftmargin=2em]
        \item $Y_{\text{continuous},j} \sim \mathcal{N}(-0.83, \sqrt{0.25})$
    \end{itemize}
    
    \item \textbf{Non-rare scenarios}:
    \begin{itemize}[leftmargin=2em]
        \item $Y_{\text{continuous},j} \sim \mathcal{N}(-0.095, \sqrt{0.25})$
    \end{itemize}
    
    \item \textbf{Binary outcome}:
    \begin{itemize}[leftmargin=2em]
        \item $Y_{\text{binary},j} = \begin{cases} +1 & \text{if } Y_{\text{continuous},j} > 0 \\ -1 & \text{otherwise} \end{cases}$
    \end{itemize}
\end{itemize}

\paragraph{Step 2: Generate visits and note variable}
\begin{itemize}[leftmargin=*]
    \item $H_j \sim \max(\text{Poisson}(2), 1)$ visits per person
    \item $S_{\text{note},j} \sim 1 + \mathcal{N}(0, 0.001^2)$ (noisy note variable for MAP)
\end{itemize}

\paragraph{Step 3: Generate ICD mentions}
\begin{itemize}[leftmargin=*]
    \item \textbf{Informative scenarios}:
    \begin{itemize}[leftmargin=2em]
        \item If $Y_{\text{binary},j} = +1$ (case):
        \begin{itemize}[leftmargin=2em]
            \item $\beta_{\text{ICD}} = 5.5$
            \item $S_{\text{ICD},j} = \max((\text{Gamma}(5.5, 1) \times H_j^{0.3} - 1), 1)$
        \end{itemize}
        \item If $Y_{\text{binary},j} = -1$ (control):
        \begin{itemize}[leftmargin=2em]
            \item $\beta_{\text{ICD}} = 0.71$
            \item $S_{\text{ICD},j} = \max((\text{Gamma}(0.71, 1) \times H_j^{0.3} - 1), 0)$
        \end{itemize}
    \end{itemize}
    
    \item \textbf{Non-informative scenarios}:
    \begin{itemize}[leftmargin=2em]
        \item Both case and control use $\beta_{\text{ICD}} = 3.0$
        \item $S_{\text{ICD},j} = \max((\text{Gamma}(3.0, 1) \times H_j^{0.3} - 1), \text{minimum})$
    \end{itemize}
\end{itemize}

\paragraph{Step 4: Generate NLP mentions}
\begin{itemize}[leftmargin=*]
    \item \textbf{Informative scenarios}:
    \begin{itemize}[leftmargin=2em]
        \item If $Y_{\text{binary},j} = +1$ (case):
        \begin{itemize}[leftmargin=2em]
            \item $\beta_{\text{NLP}} = 2.1$
            \item $S_{\text{NLP},j} = \max((\text{Gamma}(2.1, 1) \times H_j^{0.25} - 1), 1)$
        \end{itemize}
        \item If $Y_{\text{binary},j} = -1$ (control):
        \begin{itemize}[leftmargin=2em]
            \item $\beta_{\text{NLP}} = 0.81$
            \item $S_{\text{NLP},j} = \max((\text{Gamma}(0.81, 1) \times H_j^{0.25} - 1), 0)$
        \end{itemize}
    \end{itemize}
    
    \item \textbf{Non-informative scenarios}:
    \begin{itemize}[leftmargin=2em]
        \item Both case and control use $\beta_{\text{NLP}} = 1.0$
        \item $S_{\text{NLP},j} = \max((\text{Gamma}(1.0, 1) \times H_j^{0.25} - 1), \text{minimum})$
    \end{itemize}
\end{itemize}

\paragraph{Step 5: Generate 150 additional NLP features}
\begin{itemize}[leftmargin=*]
    \item For each feature $k = 1, \ldots, 150$:
    \begin{itemize}[leftmargin=2em]
        \item If $Y_{\text{binary},j} = +1$ (case): $b_k = 1.25$
        \item If $Y_{\text{binary},j} = -1$ (control): $b_k = 0.9$
        \item NLP$_{k,j} = \max((\text{Gamma}(b_k, 1) \times H_j^{0.2} - 1), 0)$
    \end{itemize}
\end{itemize}

\paragraph{Step 6: Calculate final probability}
\begin{itemize}[leftmargin=*]
    \item Transform continuous outcome to probability:
    \begin{itemize}[leftmargin=2em]
        \item $q_j = \frac{1}{1+\exp(-Y_{\text{continuous},j})}$
    \end{itemize}
    
    \item Calculate log likelihoods:
    \begin{itemize}[leftmargin=2em]
        \item $\log L_{\text{case}} = \log P(S_{\text{ICD},j}|\text{Gamma}(\beta^+_{\text{ICD}}, 1)) + \log P(S_{\text{NLP},j}|\text{Gamma}(\beta^+_{\text{NLP}}, 1))$
        \item $\log L_{\text{control}} = \log P(S_{\text{ICD},j}|\text{Gamma}(\beta^-_{\text{ICD}}, 1)) + \log P(S_{\text{NLP},j}|\text{Gamma}(\beta^-_{\text{NLP}}, 1))$
    \end{itemize}
    
    \item Update probability:
    \begin{itemize}[leftmargin=2em]
        \item $\log \text{numerator} = \log(q_j) + \log L_{\text{case}}$
        \item $\log \text{denominator} = \log(q_j \times \exp(\log L_{\text{case}}) + (1-q_j) \times \exp(\log L_{\text{control}}))$
        \item $p_j = \exp(\log \text{numerator} - \log \text{denominator})$
    \end{itemize}
\end{itemize}

\subsubsection{Method 3: Complex Data Generation}

\paragraph{Overview}
This most realistic approach incorporates demographics, multiple encounters, and dependent feature generation that mirrors real clinical workflows.

\paragraph{Parameters:}
\begin{itemize}[leftmargin=*]
    \item $n = 10,000$ individuals
    \item Test set size = 200
    \item Scenarios: rare\_info, rare\_noninfo, nonrare\_info, nonrare\_noninfo
\end{itemize}

\paragraph{Step 1: Generate demographic variables X and baseline probability Y as function of X}

\textbf{Generate demographic variables:}
\begin{itemize}[leftmargin=*]
    \item $X = (\text{age}, \text{sex}, \text{race}, \text{medical history/exposures})$
    
    \item \textbf{Age}:
    \begin{itemize}[leftmargin=2em]
        \item Age$_j \sim \mathcal{N}(40.7, 22.6^2)$, truncated to $[0, 120]$
    \end{itemize}
    
    \item \textbf{Sex}:
    \begin{itemize}[leftmargin=2em]
        \item Sex$_j \sim \text{Bernoulli}(0.51)$ where $1 =$ Female
    \end{itemize}
    
    \item \textbf{Race}:
    \begin{itemize}[leftmargin=2em]
        \item Race$_j \sim$ Categorical with probabilities:
        \begin{itemize}
            \item White: 0.61, Black: 0.073, Asian: 0.13
            \item Native American: 0.017, Pacific Islander: 0.015
            \item Other: 0.034, Unknown: 0.12
        \end{itemize}
    \end{itemize}
    
    \item \textbf{Medical history (7 exposure types)}:
    \begin{itemize}[leftmargin=2em]
        \item Number selected: $K_j \sim \text{Poisson}(5)$
        \item Each exposure: Exposure$_{jk} \sim \text{Bernoulli}(0.30)$
    \end{itemize}
\end{itemize}

\textbf{Baseline probability generation:}
\begin{itemize}[leftmargin=*]
    \item $\text{Logit}(P(Y_j = 1|X_j)) = X_j\beta_Y$
    \item $X_j = [1, \text{Age}_j, \text{Sex}_j, \text{Race}_j, \text{Exposure}_{j1}, \ldots, \text{Exposure}_{j7}]$
    
    \item \textbf{Rare disease scenarios}:
    \begin{itemize}[leftmargin=2em]
        \item $\beta_Y = [-5.7, 0.05, 0.03, 0.1, \underbrace{0.2, \ldots, 0.2}_{\text{7 values}}]$
    \end{itemize}
    
    \item \textbf{Non-rare disease scenarios}:
    \begin{itemize}[leftmargin=2em]
        \item $\beta_Y = [-2.8, 0.05, 0.03, 0.1, \underbrace{0.2, \ldots, 0.2}_{\text{7 values}}]$
    \end{itemize}
    
    \item \textbf{Final initial probability}:
    \begin{itemize}[leftmargin=2em]
        \item $P_{\text{initial},j} = \frac{1}{1+\exp(-X_j\beta_Y)}$
    \end{itemize}
\end{itemize}

\textbf{True outcome generation:}
\begin{itemize}[leftmargin=*]
    \item $Y_j \sim \text{Bernoulli}(P_{\text{initial},j})$
\end{itemize}

\paragraph{Step 2: Generate a number of notes $n_i$ for each person j within the window}
\begin{itemize}[leftmargin=*]
    \item $H_j \sim \max(\text{Poisson}(5), 1)$ visits/notes per person
    \item Expand data to encounter level: $n_{ij} = H_j$ encounters for person $j$
    \item $S_{\text{note},j} \sim 1 + \mathcal{N}(0, 0.001^2)$ (noisy note variable)
\end{itemize}

\paragraph{Step 3: For each note/encounter i, silver labels generate as functions of X and Y}

\textbf{Text string binary count generation:}
\begin{itemize}[leftmargin=*]
    \item $T_{ij}$: the more abundant text string (i.e., ``anaphylaxis'')
    \begin{itemize}[leftmargin=2em]
        \item $T_{ij} \sim \text{Ber}(P_T)$
        \item $\text{Logit}(P_T) = X_j\beta_T + \text{offset}$
        \item \textbf{Informative scenarios}:
        \begin{itemize}
            \item $\beta_T = [-2, 0.03, -0.2, 0.1, 0.5, -0.5, 0.8, -0.5, 0.00, 0.2, 0.5]$
            \item offset $= 0.4$ (case) or $0.1$ (control)
        \end{itemize}
        \item \textbf{Non-informative scenarios}:
        \begin{itemize}
            \item $\beta_T = [0, 0, 0, \ldots, 0]$ (all zeros)
            \item offset $= 0.1$ (both case and control)
        \end{itemize}
    \end{itemize}
    
    \item $S_{ij}$: the less abundant text string (i.e., ``anaphylactic reaction'')
    \begin{itemize}[leftmargin=2em]
        \item $S_{ij} \sim \text{Ber}(P_S)$
        \item $\text{Logit}(P_S) = X_j\beta_S + \text{offset}$
        \item \textbf{Informative scenarios}:
        \begin{itemize}
            \item $\beta_S = [-2.5, 0.02, -0.1, 0.5, -0.02, 0.00, -0.3, 0.03, 0.00, 0.7, -0.3]$
            \item offset $= 0.4$ (case) or $0.1$ (control)
        \end{itemize}
        \item \textbf{Non-informative scenarios}:
        \begin{itemize}
            \item $\beta_S = [0, 0, 0, \ldots, 0]$ (all zeros)
            \item offset $= 0.1$ (both case and control)
        \end{itemize}
    \end{itemize}
\end{itemize}

\textbf{NLP\_Silver ($C1_{ij}$) \& selected NLP ($C2_{ij} - C11_{ij}$) variables count generation:}
\begin{itemize}[leftmargin=*]
    \item \textbf{NLP Silver binary indicator}:
    \begin{itemize}[leftmargin=2em]
        \item $C1_{ij} = I(T_{ij} = 1 \text{ OR } S_{ij} = 1)$
        \item $C1_{ij} \sim \text{Ber}(P_T + P_S)$
    \end{itemize}
    
    \item \textbf{NLP counts (non-binary) when $C1_{ij} = 1$}:
    \begin{itemize}[leftmargin=2em]
        \item $\text{Logit}(p_j) = \text{random\_intercept}_j + \text{base\_prob} + P_1 \text{ or } P_0$
        \item random\_intercept$_j \sim \mathcal{N}(0, 0.5^2)$
        \item base\_prob $= 1.5$
        \item \textbf{Informative scenarios}:
        \begin{itemize}
            \item $P_1$ (case offset) $= 0.3$
            \item $P_0$ (control offset) $= 0.1$
        \end{itemize}
        \item \textbf{Non-informative scenarios}:
        \begin{itemize}
            \item $P_1$ (case offset) $= 0.1$
            \item $P_0$ (control offset) $= 0.1$
        \end{itemize}
        \item \textbf{Count generation}:
        \begin{itemize}
            \item NLP$1_{ij} = \lceil \text{Poisson}(p_j) \rceil$ if $C1_{ij} = 1$
            \item NLP$1_{ij} = 0$ if $C1_{ij} = 0$
        \end{itemize}
    \end{itemize}
    
    \item \textbf{$C2_{ij}, C3_{ij}, \ldots, C11_{ij}$ (10 selected NLP variables from 150 total)}:
    \begin{itemize}[leftmargin=2em]
        \item Randomly select 10 NLP features from 150 total generated features
        \item For each selected feature $k$:
        \begin{itemize}
            \item $\text{Logit}(p_j^{(k)}) = \text{random\_intercept}_j + \text{base\_prob} + P_1 \text{ or } P_0$
            \item Same random intercept, base probability, and offset structure as NLP silver
            \item SelectedNLP$_{k,ij} = \lceil \text{Poisson}(p_j^{(k)}) \rceil$
        \end{itemize}
    \end{itemize}
\end{itemize}

\textbf{ICD\_Code binary count generation:}
\begin{itemize}[leftmargin=*]
    \item $D_{ij} \sim \text{Ber}(P_D)$
    \item $\text{Logit}(P_D) = \beta_0 + \sum_{k=1}^{10} \beta_k \cdot \text{SelectedNLP}_{k,ij} + \beta_{11} \cdot \text{NLP\_Silver\_Count}_{ij} + \text{offset}$
    \item $X = (\text{intercept}, 10 \text{ selected NLP counts}, \text{NLP\_Silver count})$
    
    \item \textbf{Coefficients}:
    \begin{itemize}[leftmargin=2em]
        \item $\beta_0 = -0.5$ (intercept)
        \item $\beta_k \sim \text{Uniform}(-0.2, 0.2)$ for $k = 1, \ldots, 10$
        \item $\beta_{11} = 0.3$ (NLP silver coefficient)
    \end{itemize}
    
    \item \textbf{Offsets}:
    \begin{itemize}[leftmargin=2em]
        \item Informative: offset $= 0.3$ (case) or $0.1$ (control)
        \item Non-informative: offset $= 0.1$ (both)
    \end{itemize}
\end{itemize}

\paragraph{Step 4: Sum all counts across $n_i$ notes for individual j}
\begin{itemize}[leftmargin=*]
    \item \textbf{Text strings}:
    \begin{itemize}[leftmargin=2em]
        \item $T_j = \sum T_{ij}$, $S_j = \sum S_{ij}$
        \item Total\_text\_mentions$_j = T_j + S_j$
    \end{itemize}
    
    \item \textbf{NLP Silver}:
    \begin{itemize}[leftmargin=2em]
        \item NLP\_Silver (binary): $C1_j = \sum C1_{ij}$
        \item $S_{\text{NLP},j} = \sum \text{NLP}1_{ij}$
    \end{itemize}
    
    \item \textbf{ICD mention}:
    \begin{itemize}[leftmargin=2em]
        \item $S_{\text{ICD},j} = \sum D_{ij}$
    \end{itemize}
    
    \item \textbf{Selected NLP counts}:
    \begin{itemize}[leftmargin=2em]
        \item Total\_SelectedNLP$_{k,j} = \sum \text{SelectedNLP}_{k,ij}$ for $k = 1, \ldots, 10$
    \end{itemize}
\end{itemize}

\paragraph{Step 5: Update initial probability to get true probability}

\begin{itemize}[leftmargin=*]
    \item \textbf{Bayes formula}:
    \begin{itemize}[leftmargin=2em]
        \item $P(Y_j = 1|Z_j, X_j) = \frac{P(Z_j|Y_j=1,X_j) \cdot P(Y_j=1|X_j) \cdot P(X_j)}{P(Z_j|Y_j=0,X_j) \cdot P(Y_j=0) \cdot P(X_j) + P(Z_j|Y_j=1,X_j) \cdot P(Y_j=1) \cdot P(X_j)}$
        \item where $Z_j =$ observed silver labels (text mentions, NLP counts, ICD codes)
    \end{itemize}
    
    \item \textbf{Likelihood $P(Z_j|Y_j, X_j)$}:
    \begin{itemize}[leftmargin=2em]
        \item $P(Z_j|Y_j, X_j) = \bar{P}_{T,j} \cdot \bar{P}_{S,j} \cdot \bar{P}_{\text{NLP},j} \cdot \bar{P}_{\text{ICD},j} + 0.001$
        \item $\bar{P} =$ mean probability across encounters
        \item Add 0.001 for numerical stability
    \end{itemize}
    
    \item \textbf{Prior $P(X_j)$}:
    \begin{itemize}[leftmargin=2em]
        \item $P(X_j) = P(\text{Age}_j) \cdot P(\text{Sex}_j) \cdot P(\text{Race}_j) \cdot P(\text{Exposure}_j) + 0.001$
        \item $P(\text{Age}_j) = \phi(\text{Age}_j; 40.7, 22.6)$ (normal density)
        \item $P(\text{Sex}_j) = 0.51$ (female) or $0.49$ (male)
        \item $P(\text{Race}_j) =$ corresponding race probability
        \item $P(\text{Exposure}_j) = 0.30$
    \end{itemize}
\end{itemize}

\paragraph{Step 6: Sample Xs and Ys to obtain marginal E[Y]}
\begin{itemize}[leftmargin=*]
    \item Sample 10,000 individuals from generated data
    \item $\hat{P}(Y = 1) = \frac{1}{10000}\sum_{k=1}^{10000} Y_k$
\end{itemize}

\paragraph{Step 7: Final probability normalization}
\begin{itemize}[leftmargin=*]
    \item Rescale $P_{\text{real},j}$ to $[0, 1]$:
    \begin{itemize}[leftmargin=2em]
        \item probability$_j = \frac{P_{\text{real},j} - \min(P_{\text{real}})}{\max(P_{\text{real}}) - \min(P_{\text{real}})}$
    \end{itemize}
\end{itemize}

\newpage

\section{Supplemental Results}

\subsection{Summary of Data Generation Parameters}

\begin{table}[H]
\centering
\caption{Method 1 (Simplified) Parameters}
\label{tab:method1_params}
\begin{tabular}{lcccc}
\toprule
\textbf{Parameter} & \textbf{Rare Info} & \textbf{Rare NonInfo} & \textbf{NonRare Info} & \textbf{NonRare NonInfo} \\
\midrule
Initial prob mean & 0.05 & 0.05 & 0.4 & 0.4 \\
Initial prob SD & 0.1 & 0.1 & 0.1 & 0.1 \\
Case ICD $\lambda$ & 8 & 5 & 8 & 5 \\
Control ICD $\lambda$ & 2 & 5 & 2 & 5 \\
Case NLP $\lambda$ & 12 & 7.5 & 12 & 7.5 \\
Control NLP $\lambda$ & 3 & 7.5 & 3 & 7.5 \\
NLP multiplier & 1.5 & 1.5 & 1.5 & 1.5 \\
\bottomrule
\end{tabular}
\end{table}

\begin{table}[H]
\centering
\caption{Method 2 (LDA) Parameters}
\label{tab:method2_params}
\begin{tabular}{lcccc}
\toprule
\textbf{Parameter} & \textbf{Rare Info} & \textbf{Rare NonInfo} & \textbf{NonRare Info} & \textbf{NonRare NonInfo} \\
\midrule
$Y$ mean & -0.83 & -0.83 & -0.095 & -0.095 \\
$Y$ SD & 0.5 & 0.5 & 0.5 & 0.5 \\
Case ICD $\beta$ & 5.5 & 3.0 & 5.5 & 3.0 \\
Control ICD $\beta$ & 0.71 & 3.0 & 0.71 & 3.0 \\
Case NLP $\beta$ & 2.1 & 1.0 & 2.1 & 1.0 \\
Control NLP $\beta$ & 0.81 & 1.0 & 0.81 & 1.0 \\
$H$ (visits) $\lambda$ & 2 & 2 & 2 & 2 \\
ICD power & 0.3 & 0.3 & 0.3 & 0.3 \\
NLP power & 0.25 & 0.25 & 0.25 & 0.25 \\
Additional NLP $b$ (case) & 1.25 & 1.25 & 1.25 & 1.25 \\
Additional NLP $b$ (control) & 0.9 & 0.9 & 0.9 & 0.9 \\
Additional NLP power & 0.2 & 0.2 & 0.2 & 0.2 \\
\bottomrule
\end{tabular}
\end{table}

\begin{table}[H]
\centering
\caption{Method 3 (Complex) Parameters}
\label{tab:method3_params}
\begin{tabular}{lcccc}
\toprule
\textbf{Parameter} & \textbf{Rare Info} & \textbf{Rare NonInfo} & \textbf{NonRare Info} & \textbf{NonRare NonInfo} \\
\midrule
$\beta_{Y,0}$ (intercept) & -5.7 & -5.7 & -2.8 & -2.8 \\
Text offset (case) & 0.4 & 0.1 & 0.4 & 0.1 \\
Text offset (control) & 0.1 & 0.1 & 0.1 & 0.1 \\
NLP offset (case) & 0.3 & 0.1 & 0.3 & 0.1 \\
NLP offset (control) & 0.1 & 0.1 & 0.1 & 0.1 \\
ICD offset (case) & 0.3 & 0.1 & 0.3 & 0.1 \\
ICD offset (control) & 0.1 & 0.1 & 0.1 & 0.1 \\
$\beta_T$ all zeros? & No & Yes & No & Yes \\
$\beta_S$ all zeros? & No & Yes & No & Yes \\
\bottomrule
\end{tabular}
\end{table}

\begin{table}[H]
\centering
\caption{Method 3 (Complex) Fixed Parameters}
\label{tab:method3_fixed_params}
\begin{tabular}{lc}
\toprule
\textbf{Fixed Parameter} & \textbf{Value} \\
\midrule
Sample size ($n$) & 10,000 \\
Test set size & 200 \\
Number of visits ($\lambda$) & 5 \\
Age mean ($\mu$) & 40.7 \\
Age SD ($\sigma$) & 22.6 \\
Female probability & 0.51 \\
Exposure probability & 0.30 \\
Number of exposure types & 7 \\
NLP base probability & 1.5 \\
Random intercept SD & 0.5 \\
ICD intercept ($\beta_0$) & -0.5 \\
ICD NLP coefficient ($\beta_{11}$) & 0.3 \\
Total NLP features & 150 \\
Selected NLP features & 10 \\
Numerical stability constant & 0.001 \\
\bottomrule
\end{tabular}
\end{table}

\newpage

\subsection{Undefined values of the true probability}\label{sec:handling_nas}

Undefined values (NAs) occur primarily due to numerical instabilities in Bayesian 
probability updates when true probabilities approach extreme values. 
Specifically, when updating initial probabilities using Bayes' rule:

\[
P(Y_j = 1 \mid \text{data}_j)
= \frac{P(\text{data}_j \mid Y_j = 1)\, P(Y_j = 1)}
{P(\text{data}_j \mid Y_j = 1)\, P(Y_j = 1) \;+\; P(\text{data}_j \mid Y_j = 0)\, P(Y_j = 0)}.
\]

If the initial true probability \(P(Y_j = 1) \approx 0\) (individual \(j\) is very 
unlikely to be a case), the numerator becomes:
\[
P(\text{data}_j \mid Y_j = 1)\cdot 0 = 0.
\]

When both the numerator and denominator approach zero simultaneously (which 
occurs when the data are also uninformative), the updated probability becomes 
undefined (\(0/0\)), resulting in \texttt{NA} true probabilities. This occurs most 
frequently in non-informative scenarios where silver labels provide little 
discriminatory power, and particularly affects individuals with true 
probabilities near 0 or 1.

We took two approaches to handling missing true probabilities for calculating mean squared error.

\paragraph{Strategy 1: Mean Imputation with NA Removal (\texttt{na.rm})}

In this approach, missing predictions are replaced with the mean of all valid predicted probabilities,
\[
\hat{p}_j^{\text{imputed}} =
\begin{cases}
\hat{p}_j, & \text{if valid},\\[4pt]
\frac{1}{|\mathcal{P}|}\sum_{i \in \mathcal{P}} \hat{p}_i, & \text{if NA},
\end{cases}
\]
where \(\mathcal{P}\) denotes individuals with non-missing predictions. After imputation, metrics such as MSE are computed only on complete pairs using \texttt{na.rm = TRUE}:
\[
\text{MSE}_{\text{na.rm}}
= \frac{1}{|\mathcal{C}|}
\sum_{j \in \mathcal{C}} (\hat{p}_j^{\text{imputed}} - p_j)^2.
\]

Because individuals with \texttt{NA} true probabilities are disproportionately likely to have missing predictions, they are excluded both from the mean calculation and, subsequently, from the evaluation set. This results in an imputed value that is biased upward and an analytic sample enriched for moderate-probability individuals, who are easier to predict. Consequently,
\[
E\!\left[\text{MSE}_{\text{na.rm}}\right]
< 
E\!\left[\text{MSE}_{\text{true population}}\right],
\]
indicating systematically optimistic performance estimates due to the removal of the hardest-to-predict cases.

\paragraph{Strategy 2: Epsilon Smoothing (Primary Method)}

To avoid exclusion while ensuring numerical stability, we instead apply a small lower bound, \(\varepsilon = 10^{-10}\), to all probabilities used in Bayesian updates:
\[
P^{*}(\cdot) = \max\{P(\cdot), \varepsilon\}.
\]
This prevents underflow and eliminates \(0/0\) update failures. Final predictions are defined as \(\hat{p}_j = \varepsilon\) when the original value is missing, and \(\hat{p}_j = \max\{\hat{p}_j, \varepsilon\}\) otherwise. All evaluation metrics are then computed across the full set of individuals,
\[
\text{MSE}_{\varepsilon} 
= \frac{1}{n}\sum_{j=1}^{n} \bigl(\hat{p}_j - \max\{p_j, \varepsilon\}\bigr)^{2},
\]
ensuring that extreme-probability cases remain included. This strategy maintains the original sample distribution, avoids the selection bias inherent in case deletion, preserves penalization for models that fail to produce stable predictions, and reflects realistic deployment where predictions must be generated for all individuals. It also provides improved numerical stability for likelihood-based computation, particularly in LDA-style algorithms.

Although missing predictions are relatively rare (generally \(<5\%\)), their concentration in non-informative scenarios and in particular algorithms (e.g., MAP-based and sureLDA-prior variants) makes the handling strategy consequential. Figure~S1 illustrates that overall MSE distributions under the two strategies appear similar; however, closer inspection reveals consistent differences. Table~SX shows that the \texttt{na.rm} approach yields systematically lower MSE values than epsilon smoothing, with correlations exceeding \(r > 0.999\) between the two sets of estimates. Across settings, \texttt{na.rm} reduces MSE by approximately 0.001--0.003, with larger discrepancies for algorithms exhibiting higher missingness (e.g., MAP: 0.8\% missing, \(\Delta\)MSE = 0.002; sureLDA\_map\_prior: 1.5\% missing, \(\Delta\)MSE = 0.008). In rare-outcome scenarios, this underestimation can reach up to 5\%. These patterns are fully consistent with the selection bias introduced by excluding the most difficult cases from the evaluation set.

\begin{figure}[htbp]
\centering
\includegraphics[width=0.9\textwidth]{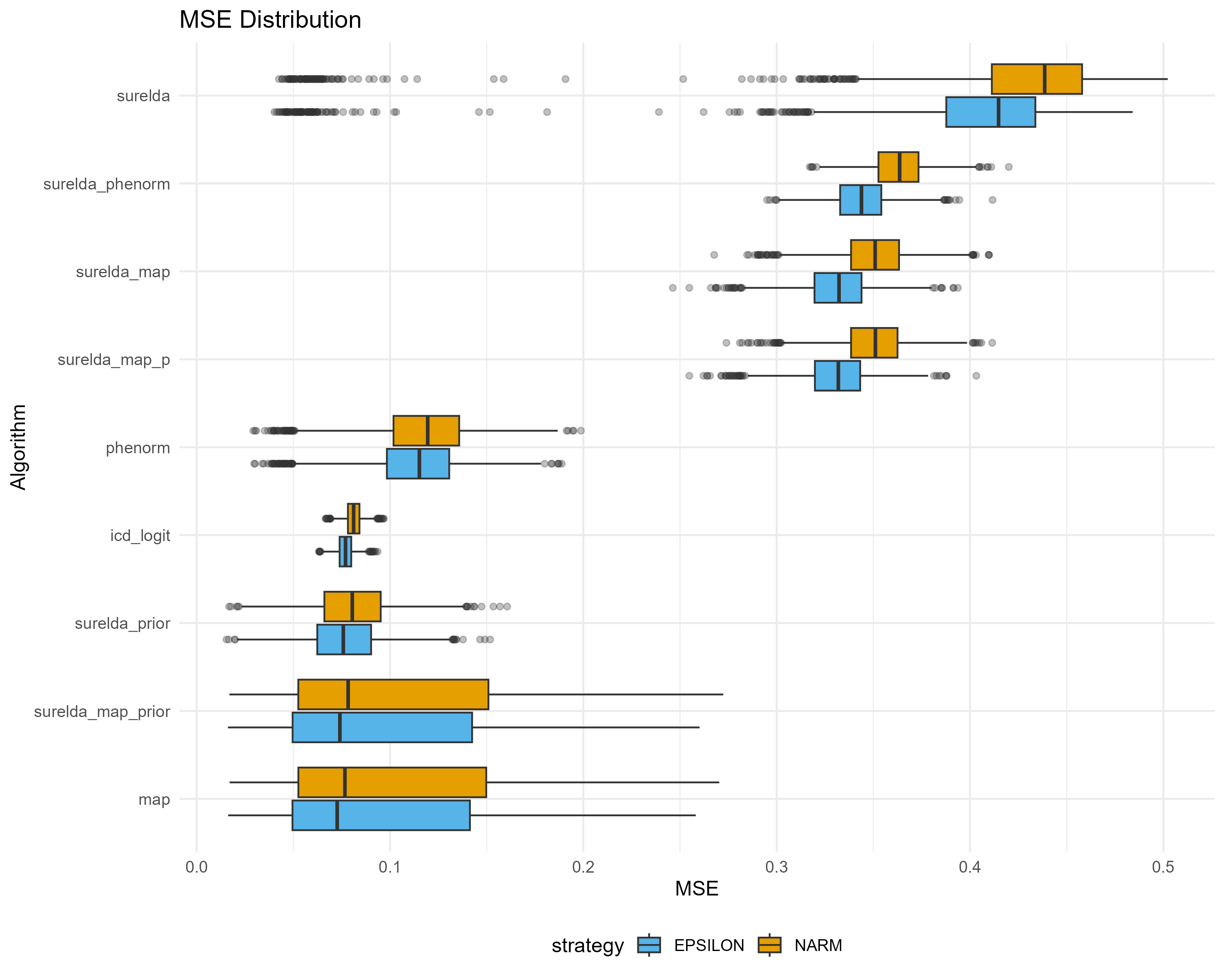}
\caption{MSE Distribution Across Different Algorithms and Strategies. Boxplots show the distribution of mean squared error (MSE) for predicted probabilities across all simulation scenarios. Error bars represent one standard deviation.}
\label{fig:mse_boxplot}
\end{figure}

\newpage

\subsection{Complete Simulation Results}

The following tables (Supplemental Tables~\ref{tab:ideal_common_info}--\ref{tab:ori_rare_noninfo}) present complete performance metrics from 2,500 simulation replicates for each scenario. Values shown are mean (standard deviation) across all replicates. Results are organized by data generation method (Simplified, LDA, Complex) and scenario (Common/Rare × Informative/Non-informative).

\begin{table}[H]
\centering
\caption{Results from the simplified data generating scenario, with a common outcome and informative silver labels.}
\label{tab:ideal_common_info}
\small
\begin{tabular}{llrrrrrrr}
\toprule
Algorithm & Set & AUC & F1 & Precision & Recall & Accuracy & Prob MSE & Prob MAE \\
\midrule
ICD Logit (v1) & Test & 0.98 & 0.91 & 0.91 & 0.91 & 0.93 & 0.05 & 0.10 \\
ICD Logit (v1) & Train & 0.98 & 0.91 & 0.92 & 0.90 & 0.93 & 0.05 & 0.10 \\
MAP (v1) & Test & 1.00 & 0.98 & 0.98 & 0.99& 0.99 & 0.04 & 0.15 \\
MAP (v1) & Train & 1.00 & 0.98 & 0.98 & 0.98 & 0.99 & 0.05 & 0.20 \\
PheNorm (v1) & Test & 1.00 & 0.99 & 0.98 & 0.99 & 0.99 & 0.01 & 0.06 \\
PheNorm (v1) & Train & 1.00 & 0.98 & 0.98 & 0.98 & 0.99 & 0.01 & 0.06 \\
SureLDA (v1) & Test & 0.86 & 0.82 & 0.73 & 0.96 & 0.82 & 0.19 & 0.22 \\
SureLDA (v1) & Train & 1.00 & 0.98 & 0.98 & 0.98 & 0.98 & 0.01 & 0.02 \\
SureLDA (v2) & Test & 0.73& 0.70 & 0.57 & 0.94 & 0.68 & 0.35 & 0.38 \\
SureLDA (v2) & Train & 1.00 & 0.96 & 0.96 & 0.97 & 0.97 & 0.04 & 0.06 \\
SureLDA (v3) & Test & 0.77 & 0.74& 0.60 & 0.96 & 0.72 & 0.29 & 0.34 \\
SureLDA (v3) & Train & 1.00 & 0.98 & 0.97 & 0.98 & 0.98 & 0.01 & 0.02 \\
SureLDA (v4) & Test & 0.72 & 0.69& 0.55 & 0.94 & 0.66 & 0.37 & 0.40 \\
SureLDA (v4) & Train & 1.00 & 0.96 & 0.95 & 0.97 & 0.97 & 0.05 & 0.07 \\
MAP (v2) & Test & 1.00 & 0.98 & 0.98 & 0.98 & 0.99 & 0.03 & 0.14 \\
MAP (v2) & Train & 1.00 & 0.98 & 0.98 & 0.98 & 0.99 & 0.05 & 0.19 \\
PheNorm (v2) & Test & 1.00 & 0.99 & 0.98 & 0.99 & 0.99 & 0.01 & 0.03 \\
PheNorm (v2) & Train & 1.00 & 0.98 & 0.98 & 0.98 & 0.99 & 0.01 & 0.03 \\
\bottomrule
\end{tabular}
\end{table}

\begin{table}[H]
\centering
\caption{Results from the simplified data generating scenario, with a common outcome and non-informative silver labels.}
\label{tab:ideal_common_noninfo}
\small
\begin{tabular}{llrrrrrrr}
\toprule
Algorithm & Set & AUC & F1 & Precision & Recall & Accuracy & Prob MSE & Prob MAE \\
\midrule
ICD Logit (v1) & Test & 0.51 & 0.43 & 0.45 & 0.53 & 0.52 & 0.01 & 0.08 \\
ICD Logit (v1) & Train & 0.50 & 0.42 & 0.41 & 0.50 & 0.51 & 0.01 & 0.08 \\
MAP (v1) & Test & 0.53 & 0.42 & 0.40 & 0.50 & 0.50 & 0.13 & 0.32 \\
MAP (v1) & Train & 0.50 & 0.42 & 0.40 & 0.49 & 0.50 & 0.09 & 0.26 \\
PheNorm (v1) & Test & 0.53 & 0.42 & 0.41 & 0.50 & 0.50 & 0.25 & 0.47 \\
PheNorm (v1) & Train & 0.50 & 0.41 & 0.41 & 0.50 & 0.51 & 0.30 & 0.52 \\
SureLDA (v1) & Test & 0.53 & 0.42 & 0.40 & 0.50 & 0.50 & 0.33 & 0.55 \\
SureLDA (v1) & Train & 0.50 & 0.42 & 0.40 & 0.50 & 0.50 & 0.10 & 0.29 \\
SureLDA (v2) & Test & 0.53 & 0.42 & 0.40 & 0.50 & 0.50 & 0.26 & 0.49 \\
SureLDA (v2) & Train & 0.50 & 0.42 & 0.40 & 0.50 & 0.50 & 0.12 & 0.31 \\
SureLDA (v3) & Test & 0.53 & 0.42 & 0.40 & 0.50 & 0.50 & 0.24 & 0.46 \\
SureLDA (v3) & Train & 0.50 & 0.43 & 0.40 & 0.50 & 0.50 & 0.25 & 0.45 \\
SureLDA (v4) & Test & 0.53 & 0.42 & 0.40 & 0.50 & 0.50 & 0.26 & 0.49 \\
SureLDA (v4) & Train & 0.50 & 0.42 & 0.40 & 0.50 & 0.50 & 0.12 & 0.31 \\
MAP (v2) & Test & 0.53 & 0.43 & 0.40 & 0.51 & 0.50 & 0.13 & 0.32 \\
MAP (v2) & Train & 0.50 & 0.42 & 0.40 & 0.50 & 0.50 & 0.09 & 0.26 \\
PheNorm (v2) & Test & 0.53 & 0.42 & 0.40 & 0.50 & 0.50 & 0.20 & 0.40 \\
PheNorm (v2) & Train & 0.50 & 0.42 & 0.40 & 0.50 & 0.50 & 0.22 & 0.43 \\
\bottomrule
\end{tabular}
\end{table}

\begin{table}[H]
\centering
\caption{Results from the simplified data generating scenario, with a rare outcome and informative silver labels.}
\label{tab:ideal_rare_info}
\small
\begin{tabular}{llrrrrrrr}
\toprule
Algorithm & Set & AUC & F1 & Precision & Recall & Accuracy & Prob MSE & Prob MAE \\
\midrule
ICD Logit (v1) & Test & 0.98 & 0.69 & 0.58 & 0.94 & 0.93 & 0.02 & 0.04 \\
ICD Logit (v1) & Train & 0.98 & 0.69 & 0.56 & 0.90 & 0.94 & 0.02 & 0.04 \\
MAP (v1) & Test & 0.99 & 0.88 & 0.82 & 0.99 & 0.97 & 0.12 & 0.22 \\
MAP (v1) & Train & 1.00 & 0.87 & 0.79 & 0.98 & 0.98 & 0.03 & 0.10 \\
PheNorm (v1) & Test & 1.00 & 0.95 & 0.92 & 0.99 & 0.99 & 0.17 & 0.36 \\
PheNorm (v1) & Train & 1.00 & 0.93 & 0.88 & 0.98 & 0.99 & 0.15 & 0.38 \\
SureLDA (v1) & Test & 0.70 & 0.25 & 0.16 & 0.90 & 0.57 & 0.41 & 0.43 \\
SureLDA (v1) & Train & 0.99 & 0.74 & 0.61 & 0.95 & 0.95 & 0.22 & 0.43 \\
SureLDA (v2) & Test & 0.69 & 0.23 & 0.14 & 0.89 & 0.54 & 0.44 & 0.46 \\
SureLDA (v2) & Train & 0.99 & 0.73 & 0.59 & 0.96 & 0.95 & 0.22 & 0.43 \\
SureLDA (v3) & Test & 0.68 & 0.24 & 0.16 & 0.78 & 0.59 & 0.47 & 0.49 \\
SureLDA (v3) & Train & 0.97 & 0.60 & 0.46 & 0.89 & 0.92 & 0.21 & 0.43 \\
SureLDA (v4) & Test & 0.69 & 0.23 & 0.14 & 0.88 & 0.55 & 0.44 & 0.46 \\
SureLDA (v4) & Train & 0.99 & 0.73 & 0.59 & 0.96 & 0.95 & 0.22 & 0.43 \\
MAP (v2) & Test & 0.99 & 0.88 & 0.82 & 0.99 & 0.97 & 0.10 & 0.20 \\
MAP (v2) & Train & 1.00 & 0.86 & 0.77 & 0.98 & 0.98 & 0.02 & 0.09 \\
PheNorm (v2) & Test & 1.00 & 0.93 & 0.89 & 0.99 & 0.99 & 0.03 & 0.11 \\
PheNorm (v2) & Train & 1.00 & 0.88 & 0.81 & 0.98 & 0.98 & 0.02 & 0.13 \\
\bottomrule
\end{tabular}
\end{table}

\begin{table}[H]
\centering
\caption{Results from the simplified data generating scenario, with a rare outcome and non-informative silver labels.}
\label{tab:ideal_rare_noninfo}
\small
\begin{tabular}{llrrrrrrr}
\toprule
Algorithm & Set & AUC & F1 & Precision & Recall & Accuracy & Prob MSE & Prob MAE \\
\midrule
ICD Logit (v1) & Test & 0.55 & 0.14 & 0.09 & 0.56 & 0.49 & 0.01 & 0.06 \\
ICD Logit (v1) & Train & 0.51 & 0.12 & 0.07 & 0.53 & 0.50 & 0.01 & 0.06 \\
MAP (v1) & Test & 0.56 & 0.13 & 0.08 & 0.50 & 0.50 & 0.43 & 0.63 \\
MAP (v1) & Train & 0.51 & 0.12 & 0.07 & 0.50 & 0.50 & 0.35 & 0.58 \\
PheNorm (v1) & Test & 0.56 & 0.13 & 0.08 & 0.50 & 0.50 & 0.66 & 0.79 \\
PheNorm (v1) & Train & 0.51 & 0.12 & 0.07 & 0.51 & 0.50 & 0.75 & 0.85 \\
SureLDA (v1) & Test & 0.56 & 0.13 & 0.08 & 0.50 & 0.50 & 0.71 & 0.77 \\
SureLDA (v1) & Train & 0.51 & 0.12 & 0.07 & 0.49 & 0.50 & 0.05 & 0.14 \\
SureLDA (v2) & Test & 0.56 & 0.13 & 0.08 & 0.50 & 0.50 & 0.46 & 0.53 \\
SureLDA (v2) & Train & 0.51 & 0.12 & 0.07 & 0.50 & 0.50 & 0.10 & 0.20 \\
SureLDA (v3) & Test & 0.56 & 0.13 & 0.08 & 0.50 & 0.50 & 0.42 & 0.50 \\
SureLDA (v3) & Train & 0.51 & 0.12 & 0.07 & 0.50 & 0.50 & 0.62 & 0.74 \\
SureLDA (v4) & Test & 0.56 & 0.13 & 0.08 & 0.50 & 0.50 & 0.46 & 0.53 \\
SureLDA (v4) & Train & 0.51 & 0.12 & 0.07 & 0.51 & 0.49 & 0.10 & 0.20 \\
MAP (v2) & Test & 0.56 & 0.13 & 0.08 & 0.50 & 0.50 & 0.43 & 0.63 \\
MAP (v2) & Train & 0.51 & 0.12 & 0.07 & 0.50 & 0.50 & 0.35 & 0.58 \\
PheNorm (v2) & Test & 0.56 & 0.13 & 0.08 & 0.50 & 0.50 & 0.55 & 0.71 \\
PheNorm (v2) & Train & 0.51 & 0.12 & 0.07 & 0.50 & 0.50 & 0.60 & 0.75 \\
\bottomrule
\end{tabular}
\end{table}

\begin{table}[H]
\centering
\caption{Results from the sureLDA data generating scenario, with a common outcome and informative silver labels.}
\label{tab:lda_common_info}
\small
\begin{tabular}{llrrrrrrr}
\toprule
Algorithm & Set & AUC & F1 & Precision & Recall & Accuracy & Prob MSE & Prob MAE \\
\midrule
ICD Logit (v1) & Test & 0.99 & 0.94 & 0.91 & 0.98 & 0.95 & 0.02 & 0.05 \\
ICD Logit (v1) & Train & 0.99 & 0.93 & 0.89 & 0.97 & 0.94 & 0.02 & 0.05 \\
MAP (v1) & Test & 0.96 & 0.92 & 0.88 & 0.97 & 0.93 & 0.10 & 0.19 \\
MAP (v1) & Train & 0.95 & 0.91 & 0.85 & 0.98 & 0.91 & 0.11 & 0.22 \\
PheNorm (v1) & Test & 1.00 & 0.98 & 0.98 & 0.99 & 0.99 & 0.03 & 0.07 \\
PheNorm (v1) & Train & 1.00 & 0.98 & 0.98 & 0.98 & 0.98 & 0.03 & 0.07 \\
SureLDA (v1) & Test & 0.99 & 0.96 & 0.94 & 0.99 & 0.97 & 0.09 & 0.13 \\
SureLDA (v1) & Train & 0.99 & 0.96 & 0.94 & 0.99 & 0.97 & 0.09 & 0.11 \\
SureLDA (v2) & Test & 0.99 & 0.95 & 0.93 & 0.97 & 0.96 & 0.02 & 0.04 \\
SureLDA (v2) & Train & 0.99 & 0.94 & 0.92 & 0.95 & 0.95 & 0.02 & 0.04 \\
SureLDA (v3) & Test & 0.99 & 0.96 & 0.93 & 0.98 & 0.96 & 0.13 & 0.18 \\
SureLDA (v3) & Train & 0.99 & 0.96 & 0.94 & 0.99 & 0.97 & 0.03 & 0.05 \\
SureLDA (v4) & Test & 0.99 & 0.95 & 0.93 & 0.97 & 0.96 & 0.03 & 0.05 \\
SureLDA (v4) & Train & 0.99 & 0.93 & 0.91 & 0.96 & 0.94 & 0.02 & 0.04 \\
MAP (v2) & Test & 0.95 & 0.92 & 0.87 & 0.98 & 0.92 & 0.10 & 0.20 \\
MAP (v2) & Train & 0.97 & 0.93 & 0.89 & 0.98 & 0.94 & 0.09 & 0.19 \\
PheNorm (v2) & Test & 1.00 & 0.97 & 0.96 & 0.99 & 0.98 & 0.02 & 0.08 \\
PheNorm (v2) & Train & 1.00 & 0.98 & 0.97 & 0.99 & 0.98 & 0.02 & 0.06 \\
\bottomrule
\end{tabular}
\end{table}

\begin{table}[H]
\centering
\caption{Results from the sureLDA data generating scenario, with a common outcome and non-informative silver labels.}
\label{tab:lda___common_noninformative}
\small
\begin{tabular}{llrrrrrrr}
\toprule
Algorithm & Set & AUC & F1 & Precision & Recall & Accuracy & Prob MSE & Prob MAE \\
\midrule
ICD Logit (v1) & Train & 0.50 & 0.54 & 0.45 & 0.73 & 0.53 & 0.02 & 0.12 \\
ICD Logit (v1) & Test & 0.53 & 0.51 & 0.42 & 0.70 & 0.51 & 0.02 & 0.12 \\
MAP (v1) & Train & 0.56 & 0.43 & 0.34 & 0.59 & 0.43 & 0.06 & 0.19 \\
MAP (v1) & Test & 0.57 & 0.35 & 0.30 & 0.43 & 0.39 & 0.09 & 0.26 \\
PheNorm (v1) & Train & 0.86 & 0.84 & 0.73 & 0.99 & 0.84 & 0.07 & 0.22 \\
PheNorm (v1) & Test & 0.80 & 0.79 & 0.68 & 0.95 & 0.79 & 0.07 & 0.22 \\
SureLDA (v1) & Train & 0.63 & 0.68 & 0.55 & 0.90 & 0.64 & 0.26 & 0.48 \\
SureLDA (v1) & Test & 0.58 & 0.62 & 0.49 & 0.88 & 0.57 & 0.25 & 0.46 \\
SureLDA (v2) & Train & 0.52 & 0.36 & 0.38 & 0.35 & 0.48 & 0.20 & 0.41 \\
SureLDA (v2) & Test & 0.57 & 0.37 & 0.33 & 0.43 & 0.40 & 0.27 & 0.50 \\
SureLDA (v4) & Train & 0.52 & 0.35 & 0.37 & 0.35 & 0.48 & 0.20 & 0.41 \\
SureLDA (v4) & Test & 0.57 & 0.37 & 0.33 & 0.43 & 0.40 & 0.27 & 0.50 \\
SureLDA (v3) & Train & 0.74 & 0.70 & 0.57 & 0.89 & 0.67 & 0.06 & 0.19 \\
SureLDA (v3) & Test & 0.72 & 0.70 & 0.59 & 0.88 & 0.68 & 0.06 & 0.21 \\
MAP (v2) & Train & 0.56 & 0.41 & 0.34 & 0.57 & 0.42 & 0.06 & 0.19 \\
MAP (v2) & Test & 0.57 & 0.35 & 0.30 & 0.42 & 0.39 & 0.09 & 0.26 \\
PheNorm (v2) & Train & 0.64 & 0.65 & 0.52 & 0.87 & 0.60 & 0.15 & 0.34 \\
PheNorm (v2) & Test & 0.53 & 0.38 & 0.34 & 0.43 & 0.42 & 0.27 & 0.49 \\
\bottomrule
\end{tabular}
\end{table}

\begin{table}[H]
\centering
\caption{Results from the sureLDA data generating scenario, with a rare outcome and informative silver labels.}
\label{tab:lda_rare_info}
\small
\begin{tabular}{llrrrrrrr}
\toprule
Algorithm & Set & AUC & F1 & Precision & Recall & Accuracy & Prob MSE & Prob MAE \\
\midrule
ICD Logit (v1) & Test & 0.99 & 0.68 & 0.54 & 1.00 & 0.95 & 0.01 & 0.03 \\
ICD Logit (v1) & Train & 0.99 & 0.55 & 0.39 & 0.97 & 0.92 & 0.01 & 0.03 \\
MAP (v1) & Test & 0.97 & 0.61 & 0.47 & 0.99 & 0.93 & 0.03 & 0.06 \\
MAP (v1) & Train & 0.97 & 0.54 & 0.39 & 0.97 & 0.91 & 0.03 & 0.06 \\
PheNorm (v1) & Test & 0.99 & 0.76 & 0.63 & 1.00 & 0.97 & 0.14 & 0.37 \\
PheNorm (v1) & Train & 0.99 & 0.66 & 0.50 & 0.98 & 0.95 & 0.13 & 0.36 \\
SureLDA (v1) & Test & 0.99 & 0.77 & 0.64 & 1.00 & 0.97 & 0.21 & 0.23 \\
SureLDA (v1) & Train & 0.99 & 0.69 & 0.53 & 0.99 & 0.96 & 0.10 & 0.10 \\
SureLDA (v2) & Test & 0.99 & 0.72 & 0.59 & 1.00 & 0.96 & 0.02 & 0.02 \\
SureLDA (v2) & Train & 0.99 & 0.67 & 0.51 & 0.96 & 0.95 & 0.01 & 0.02 \\
SureLDA (v3) & Test & 0.99 & 0.78 & 0.66 & 1.00 & 0.97 & 0.20 & 0.22 \\
SureLDA (v3) & Train & 0.99 & 0.65 & 0.49 & 0.99 & 0.95 & 0.03 & 0.05 \\
SureLDA (v4) & Test & 0.99 & 0.72 & 0.59 & 1.00 & 0.96 & 0.02 & 0.02 \\
SureLDA (v4) & Train & 0.99 & 0.67 & 0.52 & 0.96 & 0.95 & 0.01 & 0.02 \\
MAP (v2) & Test & 0.97 & 0.61 & 0.47 & 0.99 & 0.93 & 0.03 & 0.06 \\
MAP (v2) & Train & 0.96 & 0.50 & 0.35 & 0.98 & 0.90 & 0.03 & 0.07 \\
PheNorm (v2) & Test & 0.99 & 0.82 & 0.71 & 1.00 & 0.98 & 0.01 & 0.03 \\
PheNorm (v2) & Train & 0.99 & 0.69 & 0.54 & 0.97 & 0.96 & 0.01 & 0.03 \\
\bottomrule
\end{tabular}
\end{table}

\begin{table}[H]
\centering
\caption{Results from the sureLDA data generating scenario, with a rare outcome and non-informative silver labels.}
\label{tab:lda___rare_noninformative}
\small
\begin{tabular}{llrrrrrrr}
\toprule
Algorithm & Set & AUC & F1 & Precision & Recall & Accuracy & Prob MSE & Prob MAE \\
\midrule
ICD Logit (v1) & Train & 0.51 & 0.10 & 0.05 & 0.73 & 0.39 & 0.08 & 0.25 \\
ICD Logit (v1) & Test & 0.57 & 0.10 & 0.05 & 0.57 & 0.46 & 0.08 & 0.25 \\
MAP (v1) & Train & 0.60 & 0.07 & 0.04 & 0.58 & 0.33 & 0.06 & 0.19 \\
MAP (v1) & Test & 0.60 & 0.08 & 0.04 & 0.38 & 0.43 & 0.10 & 0.25 \\
PheNorm (v1) & Train & 0.73 & 0.23 & 0.14 & 0.60 & 0.81 & 0.11 & 0.28 \\
PheNorm (v1) & Test & 0.71 & 0.22 & 0.13 & 0.78 & 0.67 & 0.11 & 0.28 \\
SureLDA (v1) & Train & 0.63 & 0.14 & 0.08 & 0.89 & 0.46 & 0.39 & 0.59 \\
SureLDA (v1) & Test & 0.58 & 0.12 & 0.07 & 0.69 & 0.48 & 0.40 & 0.59 \\
SureLDA (v2) & Train & 0.53 & 0.07 & 0.04 & 0.40 & 0.51 & 0.14 & 0.32 \\
SureLDA (v2) & Test & 0.59 & 0.06 & 0.03 & 0.34 & 0.43 & 0.33 & 0.52 \\
SureLDA (v4) & Train & 0.54 & 0.07 & 0.04 & 0.42 & 0.50 & 0.14 & 0.32 \\
SureLDA (v4) & Test & 0.58 & 0.06 & 0.03 & 0.35 & 0.42 & 0.33 & 0.52 \\
SureLDA (v3) & Train & 0.59 & 0.11 & 0.06 & 0.90 & 0.30 & 0.06 & 0.20 \\
SureLDA (v3) & Test & 0.68 & 0.18 & 0.11 & 0.81 & 0.59 & 0.08 & 0.23 \\
MAP (v2) & Train & 0.60 & 0.07 & 0.04 & 0.57 & 0.33 & 0.06 & 0.20 \\
MAP (v2) & Test & 0.60 & 0.08 & 0.04 & 0.37 & 0.43 & 0.10 & 0.25 \\
PheNorm (v2) & Train & 0.66 & 0.14 & 0.08 & 0.76 & 0.55 & 0.14 & 0.31 \\
PheNorm (v2) & Test & 0.57 & 0.12 & 0.06 & 0.67 & 0.49 & 0.34 & 0.54 \\
\bottomrule
\end{tabular}
\end{table}

\begin{table}[H]
\centering
\caption{Results from the complex data generating scenario, with a common outcome and informative silver labels.}
\label{tab:ori_common_info}
\small
\begin{tabular}{llrrrrrrr}
\toprule
Algorithm & Set & AUC & F1 & Precision & Recall & Accuracy & Prob MSE & Prob MAE \\
\midrule
ICD Logit (v1) & Test & 0.86 & 0.75 & 0.74 & 0.77 & 0.79 & 0.11 & 0.27 \\
ICD Logit (v1) & Train & 0.86 & 0.75 & 0.73 & 0.77 & 0.78 & 0.11 & 0.27 \\
MAP (v1) & Test & 0.85 & 0.75 & 0.75 & 0.77 & 0.79 & 0.09 & 0.24 \\
MAP (v1) & Train & 0.84 & 0.73 & 0.70 & 0.76 & 0.77 & 0.07 & 0.22 \\
PheNorm (v1) & Test & 0.89 & 0.80 & 0.80 & 0.80 & 0.83 & 0.14 & 0.31 \\
PheNorm (v1) & Train & 0.89 & 0.78 & 0.78 & 0.79 & 0.82 & 0.14 & 0.30 \\
SureLDA (v1) & Test & 0.91 & 0.82 & 0.82 & 0.83 & 0.85 & 0.26 & 0.45 \\
SureLDA (v1) & Train & 0.90 & 0.79 & 0.78 & 0.80 & 0.83 & 0.32 & 0.50 \\
SureLDA (v2) & Test & 0.93 & 0.86 & 0.82 & 0.91 & 0.88 & 0.22 & 0.40 \\
SureLDA (v2) & Train & 0.96 & 0.89 & 0.89 & 0.89 & 0.91 & 0.18 & 0.36 \\
SureLDA (v3) & Test & 0.94 & 0.87 & 0.84 & 0.92 & 0.89 & 0.22 & 0.40 \\
SureLDA (v3) & Train & 0.97 & 0.90 & 0.90 & 0.90 & 0.92 & 0.19 & 0.36 \\
SureLDA (v4) & Test & 0.93 & 0.86 & 0.82 & 0.91 & 0.88 & 0.22 & 0.40 \\
SureLDA (v4) & Train & 0.96 & 0.89 & 0.89 & 0.89 & 0.91 & 0.18 & 0.36 \\
MAP (v2) & Test & 0.85 & 0.76 & 0.75 & 0.78 & 0.79 & 0.09 & 0.25 \\
MAP (v2) & Train & 0.85 & 0.73 & 0.73 & 0.74 & 0.78 & 0.07 & 0.22 \\
PheNorm (v2) & Test & 0.96 & 0.89 & 0.90 & 0.89 & 0.91 & 0.12 & 0.28 \\
PheNorm (v2) & Train & 0.96 & 0.88 & 0.89 & 0.88 & 0.90 & 0.12 & 0.27 \\
\bottomrule
\end{tabular}
\end{table}

\begin{table}[H]
\centering
\caption{Results from the complex data generating scenario, with a common outcome and non-informative silver labels.}
\label{tab:ori_common_noninfo}
\small
\begin{tabular}{llrrrrrrr}
\toprule
Algorithm & Set & AUC & F1 & Precision & Recall & Accuracy & Prob MSE & Prob MAE \\
\midrule
ICD Logit (v1) & Test & 0.72 & 0.61 & 0.61 & 0.64 & 0.67 & 0.07 & 0.21 \\
ICD Logit (v1) & Train & 0.72 & 0.60 & 0.60 & 0.61 & 0.67 & 0.07 & 0.21 \\
MAP (v1) & Test & 0.67 & 0.60 & 0.59 & 0.66 & 0.65 & 0.16 & 0.33 \\
MAP (v1) & Train & 0.67 & 0.59 & 0.56 & 0.63 & 0.64 & 0.13 & 0.30 \\
PheNorm (v1) & Test & 0.68 & 0.59 & 0.60 & 0.62 & 0.66 & 0.25 & 0.44 \\
PheNorm (v1) & Train & 0.68 & 0.55 & 0.56 & 0.55 & 0.64 & 0.26 & 0.46 \\
SureLDA (v1) & Test & 0.72 & 0.63 & 0.62 & 0.66 & 0.69 & 0.32 & 0.51 \\
SureLDA (v1) & Train & 0.70 & 0.60 & 0.58 & 0.63 & 0.66 & 0.39 & 0.57 \\
SureLDA (v2) & Test & 0.68 & 0.62 & 0.56 & 0.72 & 0.64 & 0.24 & 0.42 \\
SureLDA (v2) & Train & 0.76 & 0.66 & 0.64 & 0.68 & 0.71 & 0.14 & 0.31 \\
SureLDA (v3) & Test & 0.67 & 0.60 & 0.55 & 0.70 & 0.63 & 0.23 & 0.41 \\
SureLDA (v3) & Train & 0.76 & 0.66 & 0.64 & 0.69 & 0.71 & 0.12 & 0.28 \\
SureLDA (v4) & Test & 0.70 & 0.63 & 0.58 & 0.71 & 0.66 & 0.24 & 0.42 \\
SureLDA (v4) & Train & 0.77 & 0.67 & 0.66 & 0.68 & 0.72 & 0.14 & 0.31 \\
MAP (v2) & Test & 0.67 & 0.61 & 0.58 & 0.67 & 0.65 & 0.15 & 0.33 \\
MAP (v2) & Train & 0.67 & 0.59 & 0.55 & 0.66 & 0.63 & 0.12 & 0.29 \\
PheNorm (v2) & Test & 0.75 & 0.66 & 0.66 & 0.67 & 0.72 & 0.15 & 0.32 \\
PheNorm (v2) & Train & 0.75 & 0.65 & 0.63 & 0.68 & 0.71 & 0.15 & 0.32 \\
\bottomrule
\end{tabular}
\end{table}

\begin{table}[H]
\centering
\caption{Results from the complex data generating scenario, with a rare outcome and informative silver labels.}
\label{tab:ori_rare_info}
\small
\begin{tabular}{llrrrrrrr}
\toprule
Algorithm & Set & AUC & F1 & Precision & Recall & Accuracy & Prob MSE & Prob MAE \\
\midrule
ICD Logit (v1) & Test & 0.86 & 0.34 & 0.23 & 0.81& 0.80 & 0.03& 0.10 \\
ICD Logit (v1) & Train & 0.86 & 0.30 & 0.19 & 0.77 & 0.79 & 0.02 & 0.10 \\
MAP (v1) & Test & 0.84& 0.31 & 0.20 & 0.88 & 0.75 & 0.29 & 0.47 \\
MAP (v1) & Train & 0.83 & 0.26 & 0.16 & 0.82 & 0.74 & 0.19 & 0.39 \\
PheNorm (v1) & Test & 0.86 & 0.41& 0.30 & 0.82 & 0.84 & 0.45 & 0.63 \\
PheNorm (v1) & Train & 0.86 & 0.32 & 0.21 & 0.72 & 0.83 & 0.50 & 0.68 \\
SureLDA (v1) & Test & 0.91 & 0.47& 0.34 & 0.89 & 0.87 & 0.57 & 0.68 \\
SureLDA (v1) & Train & 0.92 & 0.43 & 0.30 & 0.84 & 0.87 & 0.56 & 0.66 \\
SureLDA (v2) & Test & 0.95 & 0.62 & 0.51& 0.92 & 0.92 & 0.23& 0.32 \\
SureLDA (v2) & Train & 0.97 & 0.60 & 0.46 & 0.90 & 0.93 & 0.09 & 0.23 \\
SureLDA (v3) & Test & 0.95 & 0.64 & 0.53 & 0.94 & 0.92 & 0.25 & 0.34 \\
SureLDA (v3) & Train & 0.97 & 0.62 & 0.48 & 0.90 & 0.94 & 0.06 & 0.15 \\
SureLDA (v4) & Test & 0.95 & 0.62 & 0.51 & 0.92 & 0.92 & 0.23 & 0.33 \\
SureLDA (v4) & Train & 0.97 & 0.60 & 0.46 & 0.90 & 0.93 & 0.09 & 0.23 \\
MAP (v2) & Test & 0.84 & 0.33& 0.21 & 0.87 & 0.77 & 0.28 & 0.46 \\
MAP (v2) & Train & 0.84 & 0.26 & 0.16 & 0.82 & 0.74 & 0.19 & 0.39 \\
PheNorm (v2) & Test & 0.94 & 0.58& 0.46& 0.92 & 0.91 & 0.14 & 0.31 \\
PheNorm (v2) & Train & 0.95 & 0.49 & 0.35 & 0.86 & 0.90 & 0.08 & 0.24 \\
\bottomrule
\end{tabular}
\end{table}

\begin{table}[H]
\centering
\caption{Results from the complex data generating scenario, with a rare outcome and non-informative silver labels.}
\label{tab:ori_rare_noninfo}
\small
\begin{tabular}{llrrrrrrr}
\toprule
Algorithm & Set & AUC & F1 & Precision & Recall & Accuracy & Prob MSE & Prob MAE \\
\midrule
ICD Logit (v1) & Test & 0.72 & 0.22 & 0.14 & 0.69 & 0.69 & 0.01 & 0.06 \\
ICD Logit (v1) & Train & 0.72 & 0.19 & 0.11 & 0.62 & 0.70 & 0.01 & 0.06 \\
MAP (v1) & Test & 0.67 & 0.19 & 0.12 & 0.76 & 0.61 & 0.34 & 0.52 \\
MAP (v1) & Train & 0.67 & 0.16 & 0.09 & 0.69 & 0.60 & 0.26 & 0.46 \\
PheNorm (v1) & Test & 0.68 & 0.21 & 0.14 & 0.71 & 0.66 & 0.55 & 0.70 \\
PheNorm (v1) & Train & 0.68 & 0.17 & 0.10 & 0.56 & 0.69 & 0.57 & 0.72 \\
SureLDA (v1) & Test & 0.72 & 0.24 & 0.15 & 0.74 & 0.70 & 0.60 & 0.69 \\
SureLDA (v1) & Train & 0.71 & 0.18 & 0.11 & 0.64 & 0.68 & 0.66 & 0.73 \\
SureLDA (v2) & Test & 0.57 & 0.13 & 0.07 & 0.56 & 0.53 & 0.46 & 0.53 \\
SureLDA (v2) & Train & 0.67 & 0.15 & 0.09 & 0.70 & 0.57 & 0.11 & 0.20 \\
SureLDA (v3) & Test & 0.56 & 0.11 & 0.06 & 0.49 & 0.51 & 0.42 & 0.49 \\
SureLDA (v3) & Train & 0.67 & 0.16 & 0.09 & 0.71 & 0.57 & 0.08 & 0.19 \\
SureLDA (v4) & Test & 0.58 & 0.13 & 0.07 & 0.56 & 0.54 & 0.45 & 0.52 \\
SureLDA (v4) & Train & 0.67 & 0.16 & 0.09 & 0.69 & 0.59 & 0.11 & 0.19 \\
MAP (v2) & Test & 0.67 & 0.19 & 0.12 & 0.76 & 0.62 & 0.32 & 0.50 \\
MAP (v2) & Train & 0.67 & 0.16 & 0.09 & 0.70 & 0.58 & 0.27 & 0.46 \\
PheNorm (v2) & Test & 0.74 & 0.26 & 0.17 & 0.75 & 0.72 & 0.32 & 0.49 \\
PheNorm (v2) & Train & 0.74 & 0.20 & 0.11 & 0.69 & 0.69 & 0.34 & 0.51 \\
\bottomrule
\end{tabular}
\end{table}

\subsection{Data Analysis: Validation Study Comparison}

\begin{table*}[htbp]
\centering
\caption{Comparison of Patient Characteristics Across Different Phenotyping Algorithms in Real-World Validation Study.}
\label{tab:phenotyping_comparison_all}
\resizebox{\textwidth}{!}{
\begin{tabular}{lcccccccccc}
\toprule
\textbf{Characteristic} & \textbf{Gold} & \textbf{MAP-v1} & \textbf{MAP-v2} & \textbf{PheNorm-v1} & \textbf{PheNorm-v2} & \textbf{SureLDA-v2} & \textbf{SureLDA-v4} & \textbf{SureLDA-v1} & \textbf{SureLDA-v3} & \textbf{SureLDA} \\
& \textbf{Standard} & & & & & & & & & \textbf{standalone} \\
\midrule
\cellcolor{gray!10}N & \cellcolor{gray!10}145 & \cellcolor{gray!10}200 & \cellcolor{gray!10}200 & \cellcolor{gray!10}200 & \cellcolor{gray!10}200 & \cellcolor{gray!10}200 & \cellcolor{gray!10}200 & \cellcolor{gray!10}200 & \cellcolor{gray!10}200 & \cellcolor{gray!10}200 \\
Age [Mean (SD)] & 49.1 (19) & 49 (17.8) & 48.3 (18.1) & 52.7 (18.3) & 50.3 (18.6) & 47.1 (19.8) & 44.5 (19.1) & 45.2 (20.1) & 45.2 (19.7) & 52 (17.8) \\
\cellcolor{gray!10}Sex (Female) & \cellcolor{gray!10}95 (65.5\%) & \cellcolor{gray!10}138 (69\%) & \cellcolor{gray!10}136 (68\%) & \cellcolor{gray!10}136 (68\%) & \cellcolor{gray!10}141 (70.5\%) & \cellcolor{gray!10}136 (68\%) & \cellcolor{gray!10}137 (68.5\%) & \cellcolor{gray!10}126 (63\%) & \cellcolor{gray!10}138 (69\%) & \cellcolor{gray!10}145 (72.5\%) \\
\addlinespace
\multicolumn{11}{l}{\textit{Race/Ethnicity}} \\
African American & 7 (4.8\%) & 11 (5.5\%) & 15 (7.5\%) & 11 (5.5\%) & 12 (6\%) & 16 (8\%) & 12 (6\%) & 17 (8.5\%) & 19 (9.5\%) & 12 (6\%) \\
\cellcolor{gray!10}Asian & \cellcolor{gray!10}18 (12.4\%) & \cellcolor{gray!10}20 (10\%) & \cellcolor{gray!10}22 (11\%) & \cellcolor{gray!10}21 (10.5\%) & \cellcolor{gray!10}17 (8.5\%) & \cellcolor{gray!10}28 (14\%) & \cellcolor{gray!10}28 (14\%) & \cellcolor{gray!10}27 (13.5\%) & \cellcolor{gray!10}23 (11.5\%) & \cellcolor{gray!10}23 (11.5\%) \\
Hispanic/Pacific Islander & 3 (2.1\%) & 1 (0.5\%) & 5 (2.5\%) & 2 (1\%) & 1 (0.5\%) & 4 (2\%) & 3 (1.5\%) & 2 (1\%) & 1 (0.5\%) & 3 (1.5\%) \\
\cellcolor{gray!10}Native American & \cellcolor{gray!10}1 (0.7\%) & \cellcolor{gray!10}1 (0.5\%) & \cellcolor{gray!10}3 (1.5\%) & \cellcolor{gray!10}1 (0.5\%) & \cellcolor{gray!10}1 (0.5\%) & \cellcolor{gray!10}1 (0.5\%) & \cellcolor{gray!10}0 (0\%) & \cellcolor{gray!10}2 (1\%) & \cellcolor{gray!10}2 (1\%) & \cellcolor{gray!10}1 (0.5\%) \\
Other & 5 (3.4\%) & 5 (2.5\%) & 5 (2.5\%) & 8 (4\%) & 9 (4.5\%) & 4 (2\%) & 6 (3\%) & 7 (3.5\%) & 7 (3.5\%) & 4 (2\%) \\
\cellcolor{gray!10}Unknown & \cellcolor{gray!10}19 (13.1\%) & \cellcolor{gray!10}25 (12.5\%) & \cellcolor{gray!10}22 (11\%) & \cellcolor{gray!10}13 (6.5\%) & \cellcolor{gray!10}21 (10.5\%) & \cellcolor{gray!10}15 (7.5\%) & \cellcolor{gray!10}15 (7.5\%) & \cellcolor{gray!10}13 (6.5\%) & \cellcolor{gray!10}15 (7.5\%) & \cellcolor{gray!10}18 (9\%) \\
White & 92 (63.4\%) & 137 (68.5\%) & 128 (64\%) & 144 (72\%) & 139 (69.5\%) & 132 (66\%) & 136 (68\%) & 132 (66\%) & 133 (66.5\%) & 139 (69.5\%) \\
\addlinespace
\multicolumn{11}{l}{\textit{Treatment Received}} \\
\cellcolor{gray!10}Antihistamine & \cellcolor{gray!10}91 (62.8\%) & \cellcolor{gray!10}148 (74\%) & \cellcolor{gray!10}113 (56.5\%) & \cellcolor{gray!10}147 (73.5\%) & \cellcolor{gray!10}154 (77\%) & \cellcolor{gray!10}149 (74.5\%) & \cellcolor{gray!10}150 (75\%) & \cellcolor{gray!10}145 (72.5\%) & \cellcolor{gray!10}149 (74.5\%) & \cellcolor{gray!10}151 (75.5\%) \\
Epinephrine & 90 (62.1\%) & 139 (69.5\%) & 131 (65.5\%) & 137 (68.5\%) & 141 (70.5\%) & 153 (76.5\%) & 152 (76\%) & 148 (74\%) & 151 (75.5\%) & 150 (75\%) \\
\addlinespace
\multicolumn{11}{l}{\textit{Clinical Documentation Features}} \\
\cellcolor{gray!10}ICD Codes [Mean $\pm$ SD] & \cellcolor{gray!10}1.14 $\pm$ 0.91 & \cellcolor{gray!10}1.09 $\pm$ 1.32 & \cellcolor{gray!10}1.09 $\pm$ 0.79 & \cellcolor{gray!10}0.92 $\pm$ 1.11 & \cellcolor{gray!10}0.96 $\pm$ 1.14 & \cellcolor{gray!10}0.85 $\pm$ 1.03 & \cellcolor{gray!10}0.85 $\pm$ 1.02 & \cellcolor{gray!10}0.8 $\pm$ 0.89 & \cellcolor{gray!10}0.79 $\pm$ 0.9 & \cellcolor{gray!10}0.96 $\pm$ 1.32 \\
Mentions [Mean $\pm$ SD] & 3.12 $\pm$ 4.47 & 4.34 $\pm$ 9.18 & 3.52 $\pm$ 5.66 & 5.34 $\pm$ 7.71 & 5.59 $\pm$ 9.79 & 5.74 $\pm$ 9.61 & 5.48 $\pm$ 9.87 & 5.83 $\pm$ 10.01 & 5.42 $\pm$ 9.3 & 6.11 $\pm$ 9.75 \\
\cellcolor{gray!10}CUIs [Mean $\pm$ SD] & \cellcolor{gray!10}1.37 $\pm$ 1.58 & \cellcolor{gray!10}1.52 $\pm$ 2.45 & \cellcolor{gray!10}1.41 $\pm$ 1.62 & \cellcolor{gray!10}2.01 $\pm$ 2.41 & \cellcolor{gray!10}1.92 $\pm$ 2.34 & \cellcolor{gray!10}1.9 $\pm$ 2.22 & \cellcolor{gray!10}1.91 $\pm$ 2.4 & \cellcolor{gray!10}1.9 $\pm$ 2.34 & \cellcolor{gray!10}1.97 $\pm$ 2.48 & \cellcolor{gray!10}2.12 $\pm$ 2.55 \\
\bottomrule
\end{tabular}
}
\end{table*}

\section*{Computational Environment}
\medskip
\begin{figure}[H]
    \centering
    \includegraphics[width=\textwidth]{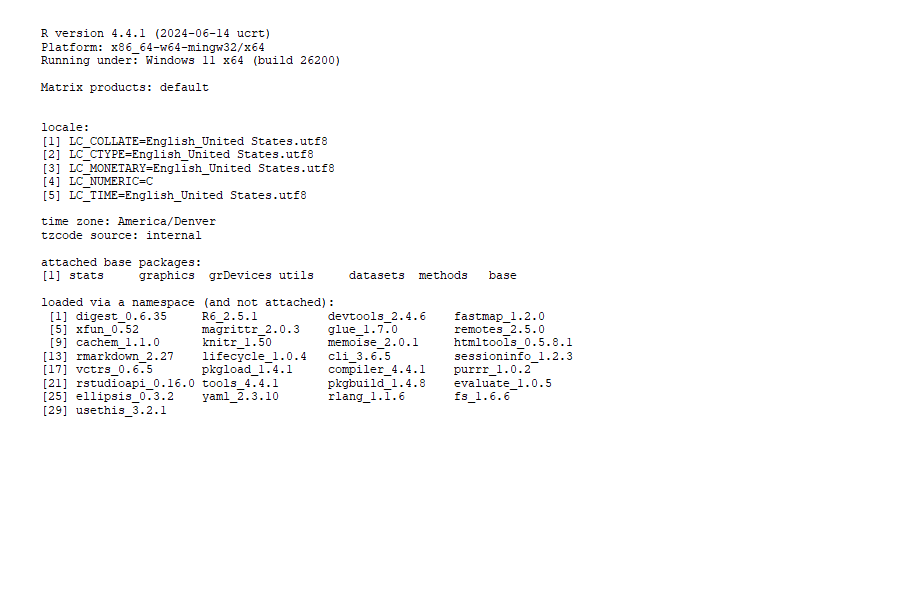}
    \caption{R session information used for all analyses.}
    \label{fig:session}
\end{figure}

\bibliographystyle{chicago}
{\footnotesize \bibliography{references}}
\clearpage

\begin{figure}[H]
    \centering
    \includegraphics[width=0.8\textwidth]{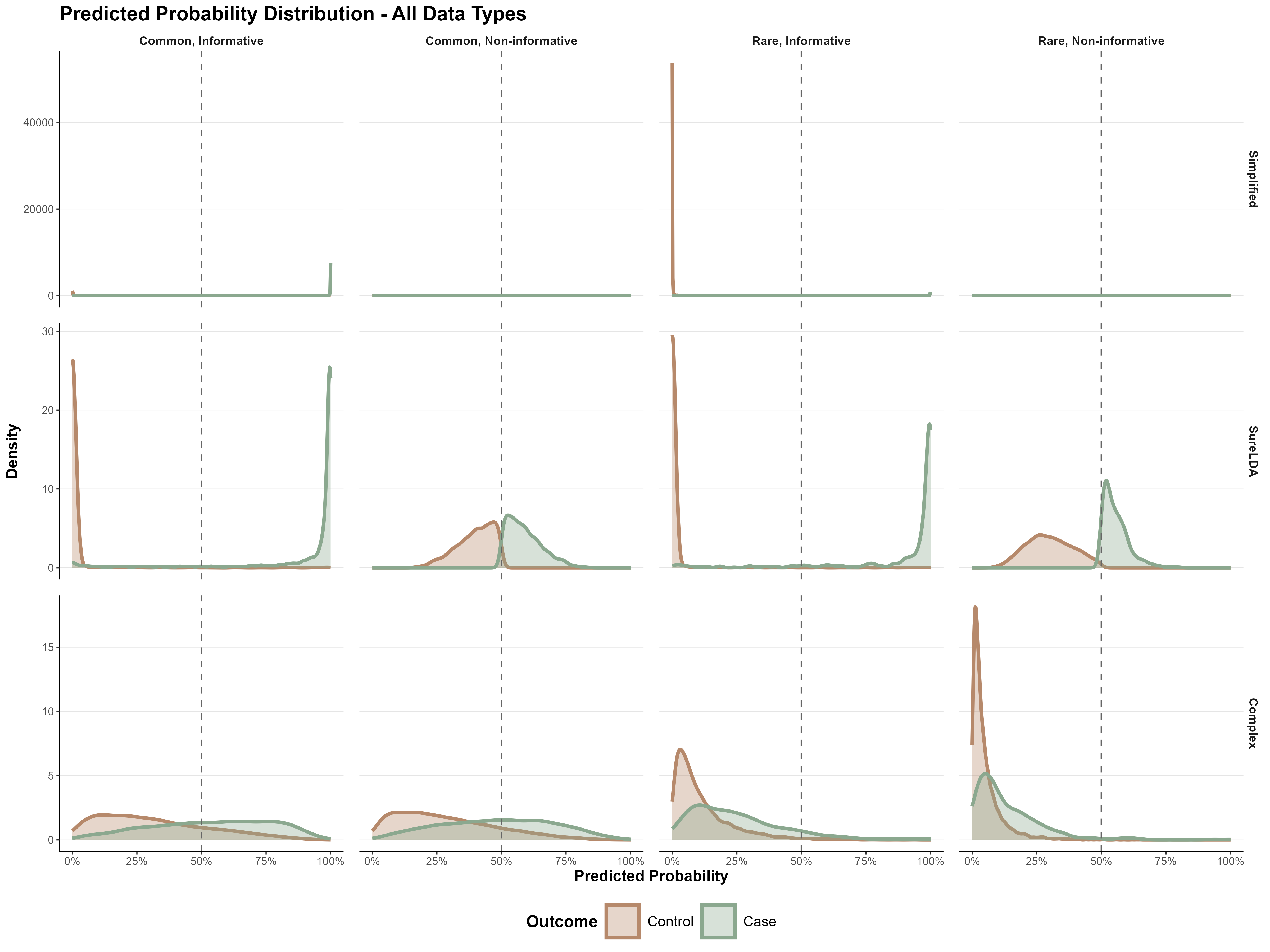}
    \caption{Distribution of outcomes and predicted probabilities in the complex data generation method.}
    \label{fig:all_data_gen}
\end{figure}

\end{document}